\documentclass[a4paper, 12pt]{article}
\usepackage{a4}
\usepackage{amsfonts}
\usepackage{amssymb}
\usepackage{cite}
\usepackage{epsf}
\usepackage{psfig}

\usepackage{graphicx}
\usepackage{subfigure}
\author{}

\newcommand{\be}{\begin{equation}}
\newcommand{\ee}{\end{equation}}
\newcommand{\bea}{\begin{eqnarray}}
\newcommand{\eea}{\end{eqnarray}}

\newcommand{\1}{{\bf 1}}
\newcommand{\3}{{\bf 3}}
\newcommand{\2}{{\bf 2}}

\newcommand{\Z}{{\mathbb Z}}
\newcommand{\Anti}{{\bf Anti}}
\newcommand{\Sym}{{\bf Sym}}

\newcommand{\N}{{\bf N}}

\newcommand{\ov}{\overline}
\def\IR{\relax{\rm I\kern-.18em R}}

\def\IP{\relax{\rm I\kern-.18em P}}
\def\inbar{\vrule height1.5ex width.4pt depth0pt}
\def\IC{\relax\,\hbox{$\inbar\kern-.3em{\rm C}$}}

\def\K3{{\bf K3}}

\def\ov{\overline}

\begin{document}

\title{
\begin{flushright} \vspace{-2cm}
{\small MPP-2005-122\\
LMU-ASC 68/05\\
\texttt{hep-th/0510170}}
\end{flushright}
\vspace{4.0cm}
One in a Billion: MSSM-like D-Brane Statistics
}
\vspace{1.0cm}
\author{\small Florian Gmeiner$^{\heartsuit,\spadesuit}$ \\
\small   Ralph~Blumenhagen$^{\heartsuit}$, Gabriele Honecker$^{\heartsuit}$, Dieter L\"ust$^{\heartsuit,\spadesuit}$,
Timo Weigand$^{\heartsuit,\spadesuit}$}
\date{}

\maketitle

\begin{center}
\emph{$^{\heartsuit}$ Max-Planck-Institut f\"ur Physik, F\"ohringer Ring 6, \\
  80805 M\"unchen, Germany } \\
\vspace{0.1cm}
\emph{$^{\spadesuit}$ Arnold-Sommerfeld-Center for Theoretical Physics, Department f\"ur Physik,
Ludwig-Maximilians-Universit\"at  M\"unchen, Theresienstra\ss e 37, 80333 M\"unchen, Germany}\\
\vspace{0.2cm}
\tt{flo, blumenha, gabriele, luest, weigand@mppmu.mpg.de}
\vspace{1.0cm}
\end{center}
\vspace{1.0cm}

\begin{abstract}
\noindent  
Continuing our recent work hep-th/0411173, we study  the statistics 
of four-dimensional, supersymmetric 
intersecting D-brane models in a toroidal orientifold 
background.
We have performed a vast computer survey  of solutions to the stringy
consistency conditions and present their statistical implications
with special emphasis on the frequency of Standard Model features.
Among the topics we discuss are the implications of the
K-theory constraints, statistical correlations among physical
quantities and an investigation of the various
statistical suppression factors arising once certain Standard Model features
are required. We estimate the frequency of an MSSM like gauge group
with  three generations to be one in a billion.

\end{abstract}

\thispagestyle{empty}
\clearpage

\tableofcontents

\section{Introduction}

The identification of MSSM-like string vacua is of obvious importance.
Methods have been developed to study compactifications
in various corners of the string  respectively M-theory moduli space.
The classes of models that  have been discussed in most detail
are certainly $E_8\times E_8$ heterotic string compactifications (e.g.~\cite{Ibanez:1987sn,Casas:1988hb,Braun:2005ux})
and Type II orientifold models with intersecting/magnetised 
D-branes (for references we refer to the most recent
review~\cite{Blumenhagen:2005mu}).

Despite the enormous effort put into this study and the unquestionable
successes in understanding the structure of these string models,
we are still lacking a single fully realistic candidate.
Models have been found which naturally  give rise to certain features
of the Standard Model, but all promising candidates have failed 
to be realistic at a certain step. Before  getting too desperate about
these shortcomings, though, one should keep in mind that 
the current search is restricted to very special corners of the overall configuration
space, namely those which are  technically under good control
such as toroidal orbifolds or Gepner model orientifolds~\cite{bw98,Brunner:2004zd,bw04,Dijkstra:2004cc}. A scan of all
possible models is still far beyond the present state of the art.
Therefore, the reason why no perfect model has been found yet might
simply be that there are too many string vacua and that the, 
say, one million models encountered so far are only
the tip of the iceberg of the enormous plethora of 
possible string vacua. 

The existence of a very large number
of string vacua has found convincing  support recently by the
study of flux vacua (for references see the review~\cite{Grana:2005jc}).
The fluxes induce a superpotential
which allows to freeze the former moduli fields related to the
size and shape of the underlying geometry. 
A rough estimate of the number of stable minima gave 
that there might exist  of the order of $10^{500}$ string vacua. 
During the last two years, this huge number  has triggered new ideas both on
fine tuning problems in particle physics and cosmology 
and on the right approach towards the problem of identifying  
realistic string vacua. 
In \cite{Douglas:2003um} it was advocated that, complementary to the model by model
search, a statistical approach to the string vacuum problem might be
worthwhile to pursue. In fact such an analysis could reveal insights
into the space of string vacua which might eventually provide some hints as to
in which corner one should look for realistic models or may at least
 give us an estimate of the abundance of Standard-like models.
We might even have to  face the prospect that such a statistical
analysis is the best we can ever do.

There exists a still growing collection of work dealing with the statistics 
of both Type II and M-theory flux compactifications.
In these studies the statistical analysis is mostly concerned with the 
gravitational (closed string) sector of the 
vacua, see e.g.~\cite{Ashok:2003gk,Denef:2004ze,Giryavets:2004zr,Dine:2004is,Misra:2004ky,Conlon:2004ds,Denef:2004cf,DeWolfe:2004ns,Dienes:2004pi,Dine:2005yq,Acharya:2005ez,Distler:2005hi,Douglas:2005hq}.\footnote{Criticism of the landscape idea is expressed in~\cite{Banks:2003es,Banks:2004xh}.}
Comparatively little is known about the statistics in the gauge sector
of the theory~\cite{Dijkstra:2004ym,Kumar:2004pv,Dijkstra:2004cc,Blumenhagen:2004xx,Kumar:2005hf,Arkani-Hamed:2005yv,Vafa:2005ui}. 
Of course these two sectors are not unrelated, as, 
like in Type II, many open string couplings also depend
on the closed string moduli and therefore on the fluxes freezing them. 
Even more drastically, the closed sector back-reacts on the open string sector
giving rise to supersymmetry breaking and induced soft-terms 
on the branes~\cite{ciu03,ggjl03,Lust:2004fi,lrs04a,Lust:2005bd}.
These are all phenomenologically very important and
string theoretically  very involved issues, which in a really complete
statistical analysis have to be taken into account. 
However, practically such a thorough analysis is beyond our understanding
of the theory and therefore has to wait until these points are better
and more generally understood.

As in our first analysis \cite{Blumenhagen:2004xx},
in this paper we
follow  a more
modest approach and try to investigate for a quite well understood
concrete example
the statistical distributions of some of the main quantities  of the 
gauge theory sector like the rank of the gauge group or the number of generations.
More concretely, we study supersymmetric intersecting D-brane models on the
$T^6\over Z_2\times Z_2$ orientifold background, which has enjoyed great interest in the past (see e.g.~\cite{Forste:2000hx,Cvetic:2001nr,Larosa:2003mz,Dudas:2005jx,Blumenhagen:2005tn} among many others).
We consider the ensemble of solutions to the tadpole cancellation conditions
using a special class of supersymmetric, so-called `factorizable',  D-branes. 

In general there will appear
a D-term potential which contains scalar  fields charged under the $U(1)$
gauge symmetry and  Fayet-Iliopoulos terms depending on one half of the
closed string moduli fields (complex structure moduli for Type IIA and
K\"ahler moduli for T-dual Type IIB models).
Therefore, at this level one expects to get a moduli space of vacua,
 parameterised by combinations of closed and open string scalars
and containing regions with different gauge symmetry and chiral
matter content. 
As a result, it is not so clear what one should actually count as
a string vacuum. Our attitude is that we are not really counting
different unconnected string moduli spaces but we are counting regions in the
moduli space with different gauge groups.
 
Turning on a charged open string modulus corresponds geometrically
to a recombination process of branes and therefore leads in general to curved
objects outside our class of flat factorizable branes.
On the other hand the inclusion of
supersymmetric three-form fluxes (in the Type IIB picture) might lead
to $\mu$-terms for some of the charged non-chiral fields and therefore
freeze some of the open string moduli.
It seems that the best we can do at the moment is to count
solutions of the tadpole cancellation and K-theory conditions in the
restricted set of supersymmetric branes one can actually describe. 
Since we are doing statistics we are confident that the distributions
we find are relatively stable against the inclusion of a larger set of branes. 

In \cite{Blumenhagen:2004xx} we have mainly used the saddle point method to derive the distributions
and correlations of various physical quantities and for eight- and six-dimensional toy models we could
confirm this technique against a brute force computer search. For
the most interesting four-dimensional models the saddle point method also became
more involved and time consuming and at the same time a brute force computer search
could not be carried out in, say, a couple of weeks. 
Moreover, in \cite{Blumenhagen:2004xx} we had not taken the additional K-theory constraints~\cite{Uranga:2000xp} into account.

In this paper we report on the results of an extensive computer search
involving several computer clusters
looking for supersymmetric models which satisfy both the 
tadpole and the K-theory constraints. Using this large set of data we will
statistically analyse the question:
Among all these models, what is the percentage  satisfying
certain Standard Model requirements, like Standard Model or
Pati-Salam gauge group, a massless hypercharge and three fermion
generations?

The very special framework of open string models on which our analysis is based immediately raises a crucial question: How generic are our results in the complete moduli space of string theory? A trustworthy answer of this question can only be given by comparing the analysis of this article to analogous statistical results in corresponding corners of the landscape. 
In view of the current status in the systematic exploration of string vacua, progress seems difficult to make. As a very first step into this direction, we attempt a comparison of our large volume statistics and the small radius Gepner models of \cite{Dijkstra:2004cc} wherever the two sets of data are compatible. In particular we distinguish between topological and geometrical observables and analyse how their distributions differ (or not).

The paper is organized as follows:
Section \ref{sec_models} reviews the well known properties of the
class of models we are considering. We give details about the tadpole,
supersymmetry and K-theory constraints that arise in general
and about the additional conditions we get from requiring Standard Model-like
properties.
In section \ref{sec_comp} we explain the implementation of the problem in
algorithmic form, suitable for a systematic computer analysis. Furthermore
we address the issue of finiteness of the number of solutions.
Section \ref{sec_stat} contains the results of
our computer search in
form of statistical distributions,
while in section \ref{sec_corr} we analyse these
distributions focusing on the correlation of variables.
In section \ref{SecCon} we summarize our conclusions and give an outlook to further
directions of research.

\section{Model building ingredients}\label{sec_models}

\subsection{Parameterisation of Type II orientifolds}
\label{SecParameter}

Orientifolds with magnetised D-branes are T-dual to orientifold models of intersecting D-branes.
Although the inclusion of three-form fluxes is better understood in the Type IIB picture, 
the general models will be presented in the Type IIA picture with intersecting D6-branes, 
where many quantities have a geometrical interpretation at least for vanishing 
three-form flux. This is the case on which we concentrate in this work. 

In the Type IIA language the orientifold action is taken to be $\Omega \bar{\sigma}$, where
$\bar{\sigma}$ is an anti-holomorphic involution on the compact six-dimensional space
preserving three-cycles $\Pi_{{\rm O}6}$.
The R-R charge of the O6-planes has to be cancelled by a suitable set of stacks of $N_a$ 
D6-branes wrapping the three-cycles $\Pi_a$ and their $\Omega \bar{\sigma}$ images $\Pi'_a$.
The R-R tadpole cancellation condition can be written simply as 
\bea
\label{RRT}
\sum_{a=1}^k  N_a\, ( \Pi_a + \Pi'_a) =
          4\,   \Pi_{{\rm O}6}.
\eea

The homology group $H_3({\cal M})$ can be decomposed into its $\Omega \bar{\sigma}$ even and odd parts,
$H_3({\cal M})=H^+_3({\cal M})\oplus H^-_3({\cal M})$. If  $\alpha_I\in H^+_3({\cal M})$ and 
$\beta_I\in H^-_3({\cal M})$ form a symplectic  basis, i.e. $\alpha_I \circ \beta_J= \delta_{IJ}$, 
$\alpha_I \circ \alpha_J= \beta_I\circ\beta_J=0$ with $I,J \in \{1,\ldots, h_{2,1}+1\}$, any cycle can be expanded as
\bea
\Pi_a = \vec{X_a} \vec{\alpha} + \vec{Y_a} \vec{\beta}, \quad 
\Pi'_a =\vec{X_a} \vec{\alpha} - \vec{Y_a} \vec{\beta}, \quad
\Pi_{{\rm O}6} = {1\over 2} \vec{L}\vec{\alpha}.
\eea
Here $\vec{X_a} \vec{\alpha}\equiv\sum_{I=1}^{h_{2,1}+1} X_a^I \alpha_I$ and
$X_a^I$, $Y_a^I$ are integer valued expansion coefficients.

The supersymmetry conditions read
\bea
\vec{Y_a}\vec{F}(U)=0, \quad
 \vec{X_a}\vec{U} >0,
\eea
where $U_I=\int_{\alpha_I} \Omega_3$ are the complex structure moduli and  $F_I=\int_{\beta_I} \Omega_3$.

The  intersection number between two cycles is given by
\bea
\label{EqIntersection}
I_{ab}\equiv \Pi_a \circ \Pi_b =\vec{X_a}\vec{Y_b}-\vec{X_b}\vec{Y_a}.
\eea
The general chiral spectrum is computed from the intersection numbers as shown in table~\ref{Tchiral}.
\begin{table}[htb]
\renewcommand{\arraystretch}{1.5}
\begin{center}
\begin{tabular}{|c|c||c|c|}
\hline
reps. & multiplicity & reps. & multiplicity \\\hline
$(\N_a,\ov{\N}_b)$ & $I_{ab}$ & $\Sym_a$ & $\frac{1}{2}(I_{aa'} - I_{a{\rm O}6})$ \\
$(\N_a,\N_b)$ & $I_{ab'}$ & $\Anti_a$ & $\frac{1}{2}(I_{aa'} + I_{a{\rm O}6})$ \\ 
\hline
\end{tabular}
\caption{\small Multiplicities of the chiral spectrum from the intersection numbers.  }
\label{Tchiral}
\end{center}
\end{table}

\subsection{Gauge anomalies and K-theory constraints}

Using table~\ref{Tchiral} and ~(\ref{EqIntersection}) one can compute that the cubic $SU(N_a)$ gauge anomalies vanish 
if the tadpole cancellation condition~(\ref{RRT}) is satisfied.

Mixed abelian anomalies on the other hand do occur. The mixed gauge anomaly, for example, is of the form
\bea
\label{EqMixedU1SUN}
{\cal A}_{U(1)_a-SU(N_b)^2} &\sim& N_a (I_{ab} + I_{ab'})C_2(N_b) \nonumber\\
&=& -2 N_a \vec{Y}_a \vec{X}_b C_2(N_b)
\eea
up to terms which vanish upon tadpole cancellation. 
$C_2(N_b)$ denotes the quadratic Casimir operator of the 
fundamental representation of $SU(N_b)$. 

A massless linear combination of abelian factors 
\bea
U(1)_{massless} = \sum_a x_a U(1)_a
\eea
exists for
\bea
\label{EqU1massless}
\sum_a x_a N_a  \vec{Y}_a =0,
\eea
as can be easily seen from~(\ref{EqMixedU1SUN}). More details can be found in~\cite{Ibanez:2001nd}.

Although the tadpole cancellation condition ensures the absence of cubic non-abelian gauge anomalies
and mixed and cubic abelian gauge anomalies are cancelled by a generalized Green-Schwarz mechanism,
there exists one further model building constraint~\cite{Uranga:2000xp} due to a $\mathbb{Z}_2$-valued conserved quantity,
the K-theory charge. This quantity can be seen by introducing a probe $Sp(2) \simeq SU(2)$ brane. If the K-theory 
charge conservation is violated, there occurs a global gauge anomaly. This anomaly is manifest as the existence 
of an odd number of chiral fermions transforming in the fundamental 
representation of $Sp(2)$~\cite{Witten:1982fp}.

\subsection{The $T^6/(\Z_2 \times \Z_2)$ model}

Factorizable three-cycles in a toroidal background are parameterised by their wrapping number $(n_i,m_i)$,
$i \in \{1,2,3\}$, along the basic one-cycles $(\pi_{2i-1}, \pi_{2i})$ of $\prod_{i=1}^3 T_i^2$. Two different shapes per two-torus $T_i^2$ 
respect the $\Omega \bar{\sigma}$ symmetry $z^i \rightarrow \bar{z}^i$ and they are parameterised by $b_i\in \{0,\frac{1}{2}\}$,
\bea
\Omega \bar{\sigma}: \pi_{2i-1} \rightarrow \pi_{2i-1} -2b_i\; \pi_{2i}, \quad\quad 
\Omega \bar{\sigma}: \pi_{2i}\rightarrow - \pi_{2i}.
\eea
It is convenient to work with the $\Omega \bar{\sigma}$ even combination
\bea
\tilde{\pi}_{2i-1} = \pi_{2i-1} - b_i \pi_{2i},
\eea
because this modification does not change the 
computation of basic cycle intersection numbers
(although only $\frac{1}{1-b_i}\,\tilde{\pi}_{2i-1}$ lies in the torus lattice).

The effective overall wrapping numbers of the three-cycles are conveniently parameterised by
the integers
\bea
\label{EqDefTildeXY}
\hat{X}^0 &=& \hat{b} n_1\, n_2\, n_3, \quad \hat{Y}_0=\hat{b}\tilde{m}_1\, \tilde{m}_2\, \tilde{m}_3, \\
\hat{X}^i &=& - \hat{b} n_i\, \tilde{m}_j\, \tilde{m}_k, \quad
\hat{Y}^i = - \hat{b} \tilde{m}_i n_j n_k, \quad i,j,k \in\{1,2,3\}\;\mbox{cyclic}, \nonumber
\eea
where $\tilde{m}_i=m_i+b_i\,n_i$ is the effective wrapping number along $\tilde{\pi}_{2i-1}$
and $\hat{b}=[(1-b_1)(1-b_2)(1-b_3)]^{-1}$.
Observe that as in~\cite{Blumenhagen:2004xx} we have included a minus sign
in the definition of $\hat{X}_i$ and $\hat{Y}_i$ for $i=1,2,3$. 
Furthermore due to the overall scaling factor $\hat{b}$, which we introduce
to obtain integer valued quantities
also for tilted tori, the intersection numbers are computed from
\bea
\label{Intersection_rescaled}
I_{ab} =\hat{b}^{-2}\left(\hat{\vec{X}}_a\hat{\vec{Y}}_b-\hat{\vec{X}}_b\hat{\vec{Y}}_a\right).
\eea

In terms of these quantities, the supersymmetry condition reads
\bea
\label{EqSUSY}
\sum_{I=0}^3 \hat{Y}^I \frac{1}{U_I} = 0, \quad \quad \sum_{I=0}^3 \hat{X}^I U_I > 0,
\eea
where the $U_I$ are defined in terms of the torus radii as 
\be
  U_0=R^{(1)}_1 R^{(2)}_1 R^{(3)}_1,\quad\quad
  U_i=R^{(i)}_1 R^{(j)}_2 R^{(k)}_2,\quad i,j,k\in\{1,2,3\}\; \mbox{cyclic}.
\ee

The tadpole cancellation condition is given by
\bea\label{EqTad}
\sum_a N_a \hat{X}_a^0 &=& 8\hat{b}, \nonumber\\
\sum_a N_a \hat{X}_a^i &=& \frac{8}{1-b_i}, \quad i \in\{1,2,3\},  
\eea
i.e. we have $\hat{\vec{L}}=L\,(\hat{b},(1-b_i)^{-1})^T$ with $L=8$.
The effective wrapping numbers can easily be rewritten in the original
unscaled wrapping numbers along the basic three-cycles via the relations
($i,j,k \in\{1,2,3\}$ cyclic)
\bea
X^0 &=& \hat{b}^{-1}\hat{X}^0 ,  \nonumber\\
X^i &=& n_im_jm_k = \hat{b}^{-1}\left(-\hat{X}^i + b_j \hat{Y}^k + b_k \hat{Y}^j + b_jb_k \hat{X}^0\right),  \nonumber\\
Y^0 &=& m_1m_2m_3 = \hat{b}^{-1}\left(\hat{Y}^0 + \sum_{i=1}^3 b_i \hat{X}^i - \sum_{i=1}^3 b_j b_k \hat{Y}^i - b_1b_2b_3 \hat{X}^0\right),   \nonumber\\
Y^i &=& m_in_jn_k =\hat{b}^{-1}\left(-\hat{Y}^i - b_i \hat{X}^0\right).
\label{EqXtildeX}
\eea
These quantities are needed to implement the coprime condition
on the sets $(n_i,m_i)$ of the wrapping numbers
per two-torus,
\bea
(Y_0)^2 = \prod_{i=1}^3 \gcd(Y_0, X_i).
\eea
However, for computational convenience we will work with the rescaled
wrapping numbers~(\ref{EqDefTildeXY}), which 
fulfill the same multiplicative relations as the original ones,
\bea
\label{EqRelstildeXY}
\hat{X}^I \hat{Y}^I &=& \hat{X}^J \hat{Y}^J  \quad {\rm for}\;{\rm all}\; I,J \in \{0,\ldots,3\}, \\
\hat{X}^I \hat{X}^J &=& - \hat{Y}^K \hat{Y}^L, \quad
\hat{X}^L (\hat{Y}^L)^2 = -\hat{X}^I \hat{X}^J\hat{X}^K, \quad  
\hat{Y}^L (\hat{X}^L)^2 = -\hat{Y}^I \hat{Y}^J\hat{Y}^K,  \nonumber
\eea
where on the second line $I,J,K,L$ are permutations of $\{0,\ldots,3\}$.

The number of chiral (anti)symmetric representations of some $U(N_a)$ factor is given in table~\ref{Tchiral}.
If one imposes for phenomenological reasons the constraint $\#({\rm Sym}_a)=0$, this leads to $\#({\rm Anti}_a)=-\frac{8}{\hat{b}^2}\hat{X}_a^0\hat{Y}_a^0$ and
one can distinguish two cases:
\begin{enumerate}
\item 
$\#({\rm Anti}_a)=\#({\rm Sym}_a)=0$, where for some permutation $(I,J,K,L)$ of $(0,\ldots,3)$ 
we have the relations on supersymmetric factorizable branes
\bea
\hat{Y}^I_a = \hat{Y}^J_a  = \hat{X}^K_a  = \hat{X}^L_a =0,&& \quad \quad
\hat{X}^I_a \hat{X}^J_a = - \hat{Y}^K_a \hat{Y}^L_a \neq 0,  \nonumber\\
\hat{Y}^K_a\hat{L}^K + \hat{Y}^L_a\hat{L}^L =0, && \quad \hat{L}^K U^K  = \hat{L}^L U^L.
\label{EqNoSym1}
\eea
These are the types of branes occurring also in  compactifications to six dimensions
as explained in~\cite{Blumenhagen:2004xx}.
\item
$\#({\rm Anti}_a)\neq 0, \#({\rm Sym}_a)=0$: in this case $\hat{X}^I_a\neq 0 $ for all  $I \in \{0,\ldots,3\}$
and the constraint on the vanishing net number of chiral symmetric representations can be rephrased as
\bea
\sum_{I=0}^3 \frac{1}{\hat{X}^I_a }\hat{L}^I = 16,
\label{EqNoSym2}
\eea
while~(\ref{EqSUSY}) has to be fulfilled.
\end{enumerate}

The D-brane configurations discussed so far support $U(N_a)$ gauge factors. In addition, there exist
four different kinds of $Sp(N_b)$ gauge factors, each of them descending from branes wrapping the $\Omega\bar{\sigma}\theta^k\omega^l$ 
($k,l\in\{0,1\}$) invariant planes, where $\theta$ and $\omega$ are the two orbifold generators.
The wrapping numbers of these branes are ($i,j,k \in\{1,2,3\}$ cyclic)
\bea
\hat{X}^{0} &=& \hat{b}^2, \;\; \hat{X}^1=\hat{X}^2=\hat{X}^3=0 
\qquad {\rm or} \qquad 
\hat{X}^{i} = \frac{\hat{b}}{1-b_i}, \;\; \hat{X}^0=\hat{X}^j=\hat{X}^k=0, \nonumber\\
\hat{Y}^I &=& 0 \quad {\rm for}\;{\rm all }\; I.
\label{EqSpbranes}
\eea

The K-theory constraints on consistent chiral spectra demand that there must be an even number of chiral fermions
transforming in the fundamental representation of any possible $Sp(2)$ factor. Inserting the wrapping numbers of all 
four $Sp(2)$ candidates into~(\ref{Intersection_rescaled}) and summing over all $U(N_a)$ factors, the K-theory constraints take the form
\bea
\label{EqKTheory}
  \sum_a N_a \hat{Y}^0_a &\in& 2\mathbb{Z}, \nonumber \\
  (1-b_j)(1-b_k) \sum_a N_a \hat{Y}^i_a &\in& 2\mathbb{Z}, \quad  i,j,k \in\{1,2,3\}\;\mbox{cyclic}.
\eea

\subsection{Standard Model realisations}\label{subsec_smtypes}
\begin{table}[htb!]
\renewcommand{\arraystretch}{1.5}
\begin{center}
\begin{tabular}{|c||c||c||c|}
\hline
\hline
particle & & mult. \\\hline
\multicolumn{3}{|c|}{$U(3)_a\times Sp(2)_b \times U(1)_c \times U(1)_d$ with $Q_Y^{(S)}$}\\\hline
$Q_L$ & $(\3,\2)_{0,0}$ & $I_{ab}$ \\\hline
$u_R$ & $(\ov{\3},1)_{-1,0}+(\ov{\3},1)_{0,-1}$ & $I_{a'c}+I_{a'd}$ \\\hline
$d_R$ & $(\ov{\3},1)_{1,0}+(\ov{\3},1)_{0,1}$ & $I_{a'c'}+I_{a'd'}$ \\
$d_R$ & $(\ov{\3}_A,1)_{0,0}$ & $\frac{1}{2}(I_{aa'} +I_{a{\rm O}6})$\\\hline
$L$ & $(1,\2)_{-1,0}+(1,\2)_{0,-1}$ & $I_{bc}+I_{bd}$ \\\hline
$e_R$ & $(\1,\1)_{2,0}$ & $\frac{1}{2}(I_{cc'} -I_{c{\rm O}6})$\\
$e_R$ & $(\1,\1)_{0,2}$ & $\frac{1}{2}(I_{dd'} -I_{d{\rm O}6})$\\
$e_R$ & $(\1,\1)_{1,1}$ & $I_{cd'}$ \\ 
\hline\hline
\multicolumn{3}{|c|}{$U(3)_a\times U(2)_b \times U(1)_c \times U(1)_d$ with $Q_Y^{(S)}$}\\\hline
$Q_L$ & $(\3,\ov{\2})_{0,0}$   & $I_{ab}$   \\
$Q_L$ & $(\3,\2)_{0,0}$   & $I_{ab'}$   \\\hline
$u_R$ &  $(\ov{\3},1)_{-1,0}+(\ov{\3},1)_{0,-1}$  & $I_{a'c}+I_{a'd}$ \\ \hline
$d_R$ &  $(\ov{\3},1)_{1,0}+(\ov{\3},1)_{0,1}$  & $I_{a'c'}+I_{a'd'}$ \\  
$d_R$ & $(\ov{\3}_A,1)_{0,0}$  &  $\frac{1}{2}(I_{aa'} +I_{a{\rm O}6})$ \\ \hline
$L$ & $(1,\2)_{-1,0}+(1,\2)_{0,-1}$ & $I_{bc}+I_{bd}$  \\  
$L$ & $(1,\ov{\2})_{-1,0}+(1,\ov{\2})_{0,-1}$ & $I_{b'c}+I_{b'd}$     \\ \hline 
$e_R$ & $(\1,\1)_{2,0}$ & $\frac{1}{2}(I_{cc'} -I_{c{\rm O}6})$\\
$e_R$ & $(\1,\1)_{0,2}$ & $\frac{1}{2}(I_{dd'} -I_{d{\rm O}6})$\\
$e_R$ & $(\1,\1)_{1,1}$ & $I_{cd'}$
\\ 
\hline
\end{tabular}
\caption{\small Realisation of quarks and leptons for various hypercharges. Part 1. }
\label{Tquarksleptons1}
\end{center}
\end{table}

\begin{table}[htb!]
\renewcommand{\arraystretch}{1.5}
\begin{center}
\begin{tabular}{|c||c||c||c|}
\hline
\hline
particle & & mult. \\\hline
\multicolumn{3}{|c|}{$U(3)_a\times U(2)_b \times U(1)_c \times U(1)_d$ with $Q_Y^{(1)}$}\\\hline
$Q_L$ & $(\3,\ov{\2})_{0,0}$   & $I_{ab}$  \\\hline
$u_R$ &  $(\ov{\3}_A,1)_{0,0}$ &  $\frac{1}{2}(I_{aa'} +I_{a{\rm O}6})$ \\\hline
$d_R$ &  $(\ov{\3},1)_{-1,0}+(\ov{\3},1)_{0,-1}$   & $I_{a'c}+I_{a'd}$ \\ 
$d_R$ &  $(\ov{\3},1)_{1,0}+(\ov{\3},1)_{0,1}$ & $I_{a'c'}+I_{a'd'}$ \\\hline
$L$ & $(1,\2)_{-1,0}+(1,\2)_{0,-1}$ & $I_{bc}+I_{bd}$ \\
$L$ & $(1,\2)_{1,0}+(1,\2)_{0,1}$ & $I_{bc'}+I_{bd'}$ \\ \hline
$e_R$ & $(\1,\ov{\1}_A)_{0,0}$  &  $-\frac{1}{2}(I_{bb'} +I_{b{\rm O}6})$   
   \\ 
\hline\hline
\multicolumn{3}{|c|}{$U(3)_a\times U(2)_b \times U(1)_c \times U(1)_d$ with $Q_Y^{(2)}$}\\\hline
$Q_L$ & $(\3,\ov{\2})_{0,0}$   & $I_{ab}$  \\\hline
$u_R$ & $(\ov{\3}_A,1)_{0,0}$  &  $\frac{1}{2}(I_{aa'} +I_{a{\rm O}6})$ \\\hline
$d_R$ &  $(\ov{\3},1)_{-1,0}$  & $I_{a'c}$ \\  
$d_R$ &  $(\ov{\3},1)_{1,0}$ & $I_{a'c'}$ \\\hline
$L$ & $(1,\ov{\2})_{0,-1}$  & $I_{b'd}$ \\ \hline
$e_R$ & $(\1,\ov{\1}_A)_{0,0}$  &  $-\frac{1}{2}(I_{bb'} +I_{b{\rm O}6})$   \\ 
$e_R$ & $(\1,\1)_{1,1}$   & $I_{cd'}$  \\
 $e_R$ & $(\1,\1)_{-1,1}$   &  $I_{c'd'}$ 
   \\ 
\hline
\end{tabular}
\caption{\small Realisation of quarks and leptons for various hypercharges. Part 2. }
\label{Tquarksleptons2}
\end{center}
\end{table}
For a Standard Model like sector of $\chi$ quark and lepton generations 
in a four stack intersecting D-brane model,  the gauge group has to contain 
one of the two following factors
\begin{enumerate}
\item 
$U(3)_a \times Sp(2)_b \times U(1)_c \times U(1)_d$ 
\item 
$U(3)_a \times U(2)_b \times U(1)_c \times U(1)_d$ 
\end{enumerate}
with $\#({\rm Sym}_a)=\#({\rm Sym}_b)=0$ in order to have no exotic symmetric chiral matter of the two non-abelian factors.
The first stack is therefore of the type~(\ref{EqNoSym1}) or~(\ref{EqNoSym2}), the second can also be of the form~(\ref{EqSpbranes}). 
In some cases, the Standard Model quantum numbers can also be realised on three stacks only at the cost of having no standard
Yukawa couplings (which are not always realistic in the four stack models either).

Up to an interchange of cycles and their $\Omega\bar{\sigma}$ images, there exist three different possible definitions of the hypercharge
for quarks and left-handed leptons:
\begin{enumerate}
\item 
The `standard' definition
$Q_Y^{(S)} = \frac{1}{6}Q_a+\frac{1}{2}Q_c +\frac{1}{2}Q_d $,
valid for both choices of an $Sp(2)_b$ or $U(2)_b$ factor. This hypercharge is massless if 
$\hat{\vec{Y}}_a + \hat{\vec{Y}}_c + \hat{\vec{Y}}_d =0$. 
\item 
The first `non-standard' definition 
$Q_Y^{(1)} = -\frac{1}{3}Q_a-\frac{1}{2}Q_b$, where right-handed up-type quarks are realised as antisymmetric 
representations of $U(3)_a$~\cite{Antoniadis:2000en,Blumenhagen:2001te}. In order for this hypercharge candidate to be massless, 
$\hat{\vec{Y}}_a +\hat{\vec{Y}}_b =0$ has to be satisfied.
\item 
The second `non-standard' definition 
$Q_Y^{(2)} = -\frac{1}{3}Q_a-\frac{1}{2}Q_b+Q_d$, where again right-handed up-type quarks are realised as antisymmetric 
representations of $U(3)_a$. $\hat{\vec{Y}}_a +\hat{\vec{Y}}_b -\hat{\vec{Y}}_d =0$ is the condition for this
abelian factor to remain massless. 
\end{enumerate}
The `standard' definition of the hypercharge on four stacks admits the realisation of all Standard Model particles as 
bifundamental representations, thereby ensuring the existence of Yukawa couplings from triangles of three intersecting branes.
This case can be reduced to a three stack model by simply replacing $\frac{1}{2}(Q_c+Q_d) \rightarrow \frac{1}{2}Q_c$.

The first `non-standard' definition also has a direct reduction to  three stacks  since $Q_c$ and $Q_d$ do not appear in the definition 
of $Q_Y^{(1)}$. On the other hand, the  second `non-standard' definition really requires four stacks of branes.

The corresponding realisations of the Standard Model quarks and leptons are given in tables~\ref{Tquarksleptons1} and~\ref{Tquarksleptons2}.
Apart from the listed particles, there exist candidates for right handed neutrinos and Higgs multiplets for each possible hypercharge.

Having $\chi_Q$ generations of quarks amounts for the first case to setting
\bea
\chi_Q = I_{ab} = I_{a'c}+I_{a'd} = I_{a'c'}+I_{a'd'} + \frac{1}{2}(I_{aa'} +I_{a{\rm O}6})
\eea
and $\chi_L$ lepton generations occur for 
\bea
\chi_L = I_{bc}+I_{bd} = \frac{1}{2}(I_{cc'} -I_{c{\rm O}6}) + \frac{1}{2}(I_{dd'} -I_{d{\rm O}6}) + I_{cd'}.
\eea
For the remaining cases, the analogous formulae can be read off from tables~\ref{Tquarksleptons1}
and~\ref{Tquarksleptons2}.

\section{Computer algorithms}\label{sec_comp}

We have defined the wrapping numbers $\hat{X}^I$ and $\hat{Y}^I$ as integer valued quantities
in order to implement the supersymmetry (\ref{EqSUSY}) and tadpole (\ref{EqTad}) conditions in a
fast computer algorithm.
From the equations we can derive the following inequalities
\be\label{EqConstr}
  0 < \sum_{I=0}^3 \hat{X}^I\, {U}_I \le \sum_{I=0}^3 \hat{L}^I\, {U}_I.
\ee

In a first step possible values for the wrapping numbers $\hat{X}^I$ and
$\hat{Y}^I$ are computed using (\ref{EqConstr}) for a fixed set of complex structures.

In a second step we use the tadpole equations (\ref{EqTad}), which after 
summation can be reduced to
\be\label{EqTadhat}
  \sum_{a}S_a=C \quad\mbox{with}\quad
  S_a=\sum_{I}N_a {U}_I\hat{X}_a^I\quad\mbox{and}\quad
  C=\sum_I\hat{L}^I {U}_I.
\ee

The algorithm to find solutions to these equations computes
all possible partitions of $C$ and factorises them
into possible values for $N_a$ and $\hat{X}_a^I$, taking only factorizations
into account which match the values generated in the first step.
The results obtained in this way have to be checked again if they satisfy 
all consistency conditions, especially the K-theory constraints.

\subsection{Number of solutions}\label{sec_numsol}
It is an important question whether or not the number of solutions is infinite, because a statistical statement can only be of significance if we are
dealing with a finite set of models.
Unfortunately we have not been able to give a complete analytic proof of finiteness
of the considered space of models, but we have good numerical hints towards this assumption.

\begin{figure}[ht]
\begin{center}
\includegraphics[width=0.9\textwidth]{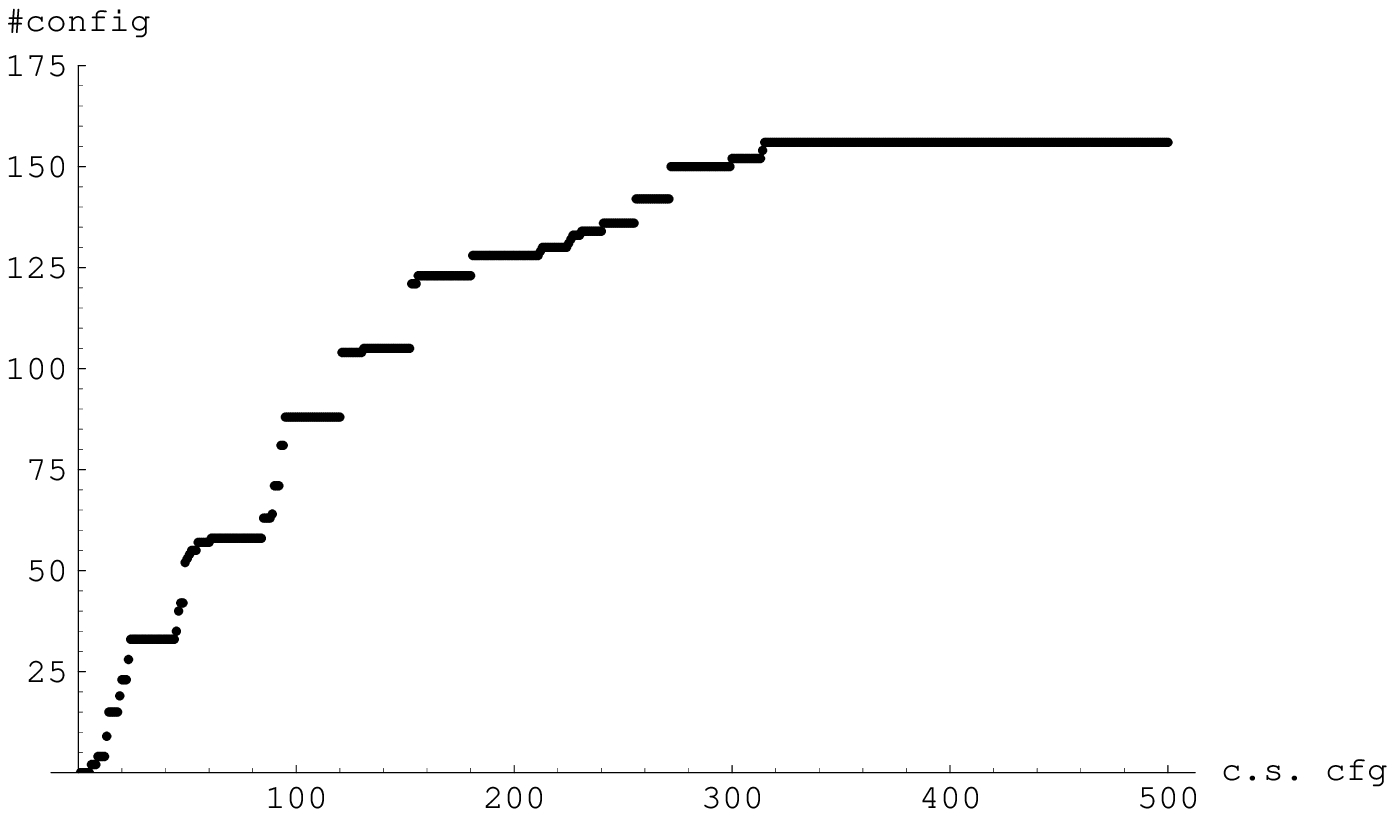}
\caption{The number of different configurations of $\hat{X}^I$ for $L=2$,
  satisfying the
  constraints. The x-axis shows combinations of complex structures 
  in an arbitrary scale.}
\label{figNumSol}
\end{center}
\begin{center}
\includegraphics[width=0.9\textwidth]{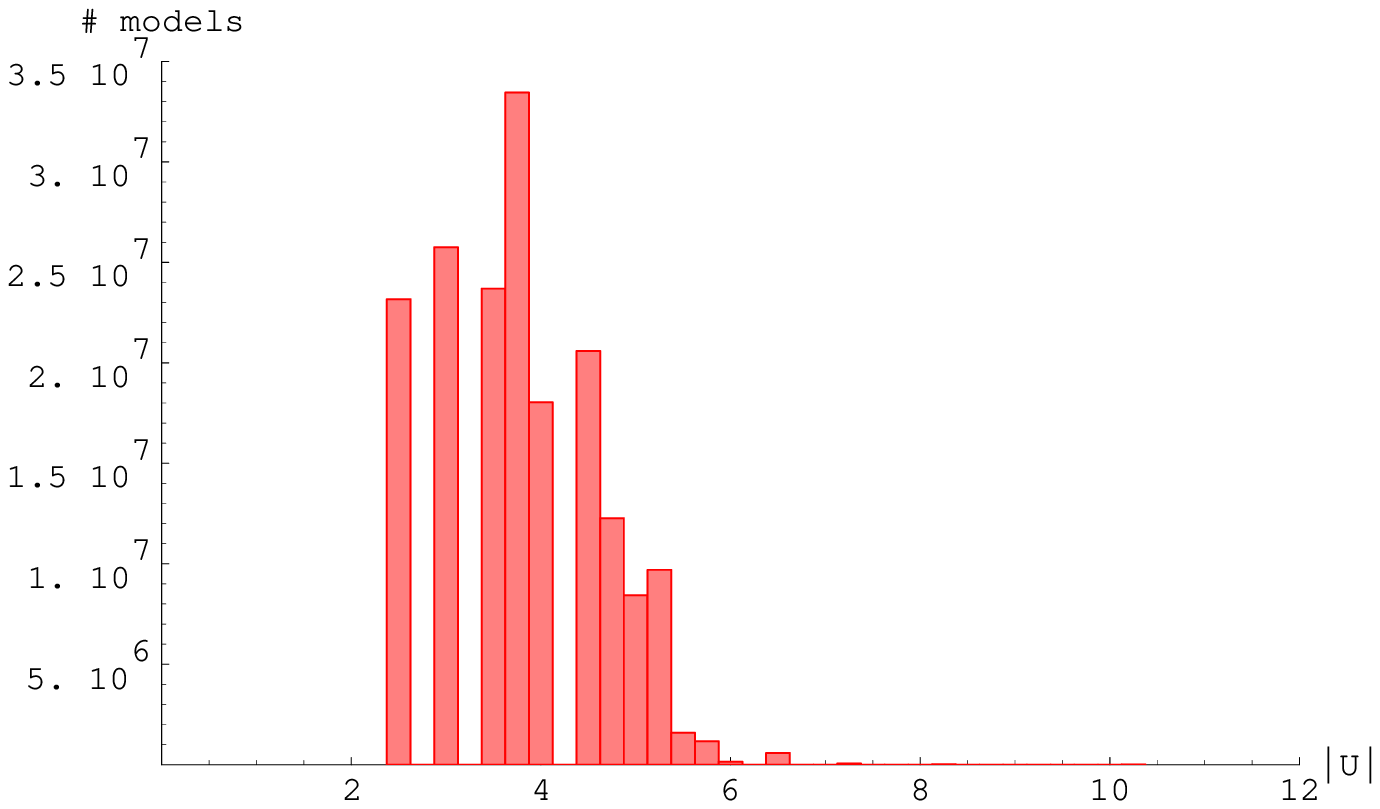}
\caption{The number of models for $L=8$, plotted against the absolute value of $\vec{U}$. The highest analysed value of $\vec{U}$ is 12.}
\label{figNumSol2}
\end{center}
\end{figure}

As explained in \cite{Blumenhagen:2004xx}, it is sufficient to analyse the
possible configurations of wrapping numbers for models in which all $\hat{X}^I$
are non-vanishing. The tadpole equations can then be shown to admit only
a finite number of supersymmetric configurations for a fixed choice of rational complex structures and fixed $L$. The difficult question is which complex
structures are compatible with the consistency conditions. For toroidal
compactifications to six dimensions it is possible to give an upper bound
for the complex structures in terms of $L$.
In the four-dimensional case, however, it is not immediately clear how to
achieve this.
Figure \ref{figNumSol} shows how the total number of mutually
different brane
configurations for $L=2$ increases and saturates,
as we include more and more combinations
of values for the complex structures $U_I$ into the set for which we construct solutions.
For this small toy value of $L$ our algorithm actually admits pushing the computations up to
those complex structures  where obviously no additional brane
solutions exist.

For the physically relevant case of $L=8$ the total number of models compared
to the absolute value  $|\vec{U}|$  of the complex structure variables scales as displayed
in figure \ref{figNumSol2}. The plot shows all complex structures we have actually been able to analyse systematically.
We find that the number of solutions falls
logarithmically for increasing values of $|\vec{U}|$.
In order to interpret this result, we observe that the complex structure moduli $U_I$ are only defined up to 
an overall rescaling by the volume modulus of the compact space. We have chosen all radii and thereby also all $U_I$ to be integer valued,  
which means that large $|\vec{U}|$ correspond to large coprime values of $R_1^{(i)}$ and $R_2^{(i)}$. This comprises on the 
one hand decompactification limits which have to  be discarded in any case for phenomenological reasons, 
but on the other hand also tori which are slightly distorted from e.g. square tori for $R_2^{(i)}/R_1^{(i)}=0.99$.

Combining the results of the two numerical tests we have reason to hope that we can
indeed make a convincing statistical statement using the analysed data.
Nevertheless, at this point it should be mentioned that we cannot fully exclude
that a large number of new solutions appears at those values for the complex structures which
we have not analysed.

\subsection{Complexity}
The main problem of the algorithm used to compute the models lies in the
fact that its complexity scales exponentially with the complex structure
parameters. Therefore we are not able to compute
up to arbitrarily high values for the $U_I$.
Although we tried our best, it may of course be possible to improve the
algorithm in many ways, but unfortunately the exponential behaviour cannot be cured unless we might have access to a quantum computer. This
is due to the fact that the problem of finding solutions to the
diophantine equations
we are considering falls in the class of NP complete problems
\cite{gareyjohnson1979}, which means that they 
cannot be reduced to problems which are solvable in
polynomial time. In fact, this is quite a severe issue since the diophantine structure of the tadpole equations encountered here is not at all exceptional but
very  generic for the topological constraints also
in other types of string  constructions.

For the total number of models analysed in this paper, which is around $1.66\times10^{8}$, we used several computer clusters built out of standard PCs
with an approximative total CPU time of roughly $4\times 10^5$ hours.
To estimate how many models could be computed in principle using a
computer
grid
equipped with contemporary technology
in a reasonable amount of time,
the exponential behaviour of the problem has to be taken into account.
Let us be optimistic and imagine that we would have a total number of
$10^5$ processors at our disposal which are
twice as fast as
the ones we have been using.
Expanding our analysis to cover a range of complex structures which is
twice as large as the one we considered  would
in a very rough estimate still take us of the order of 500 years.

Note that in principle there can be a big difference in the estimated computing
time for the two computational problems of finding all string vacua in a certain class
and of singling out special realistic ones. Apparently, for the second question,
having Standard Model features realized by just a subset or even a single stack
of branes, the search algorithm can work recursively by successively imposing
more Standard Model features. It would be interesting to know whether 
the Standard Model search can be performed by an algorithm which is polynomial.

\section{Statistical distributions}\label{sec_stat}
\subsection{Effect of the K-theory constraints}
\begin{figure}[htb!]
\begin{center}
\subfigure[Rank distribution]{\label{Fig_k_rkdist}\includegraphics[width=0.9\textwidth]{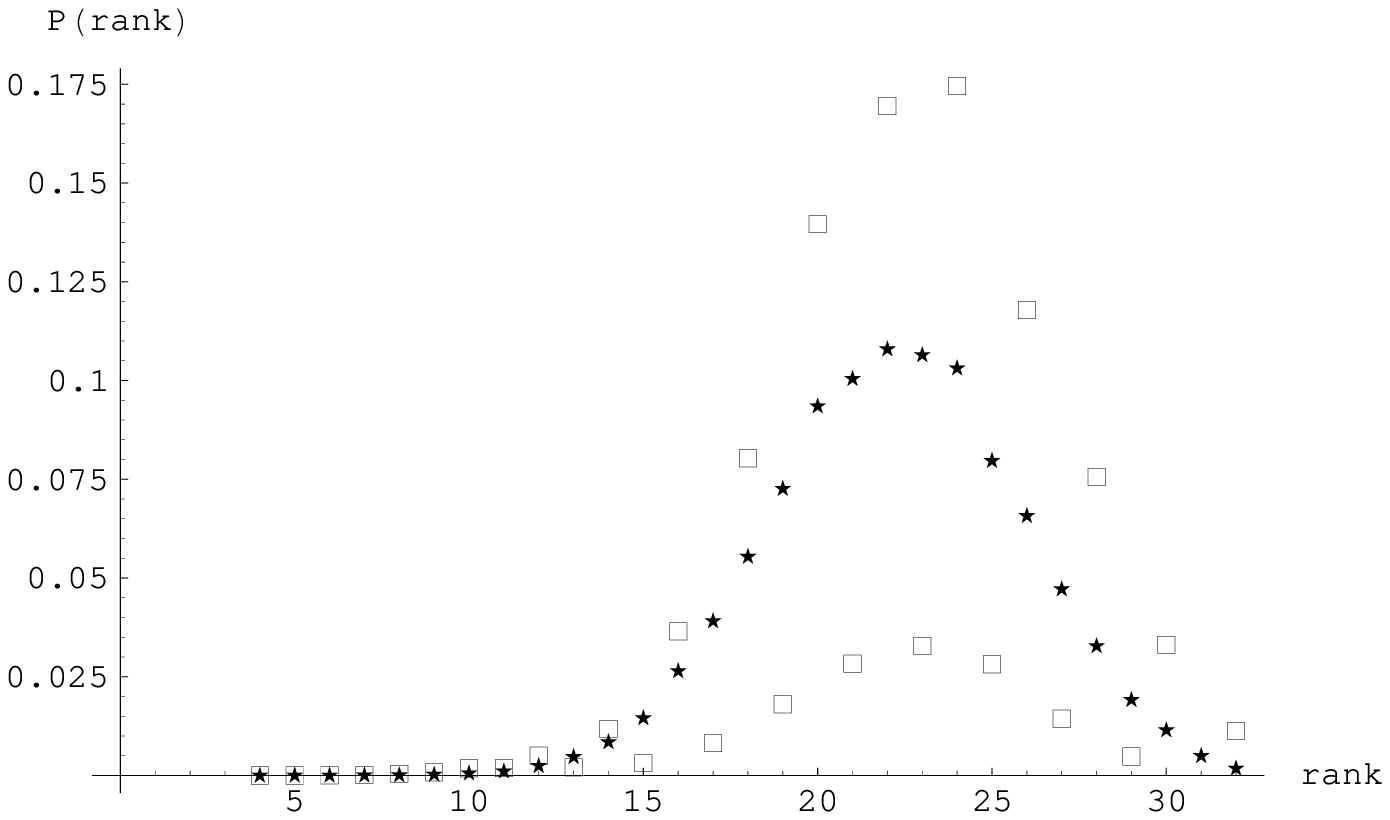}}
\subfigure[$U(M)$ distribution]{\label{Fig_k_undist}\includegraphics[width=0.9\textwidth]{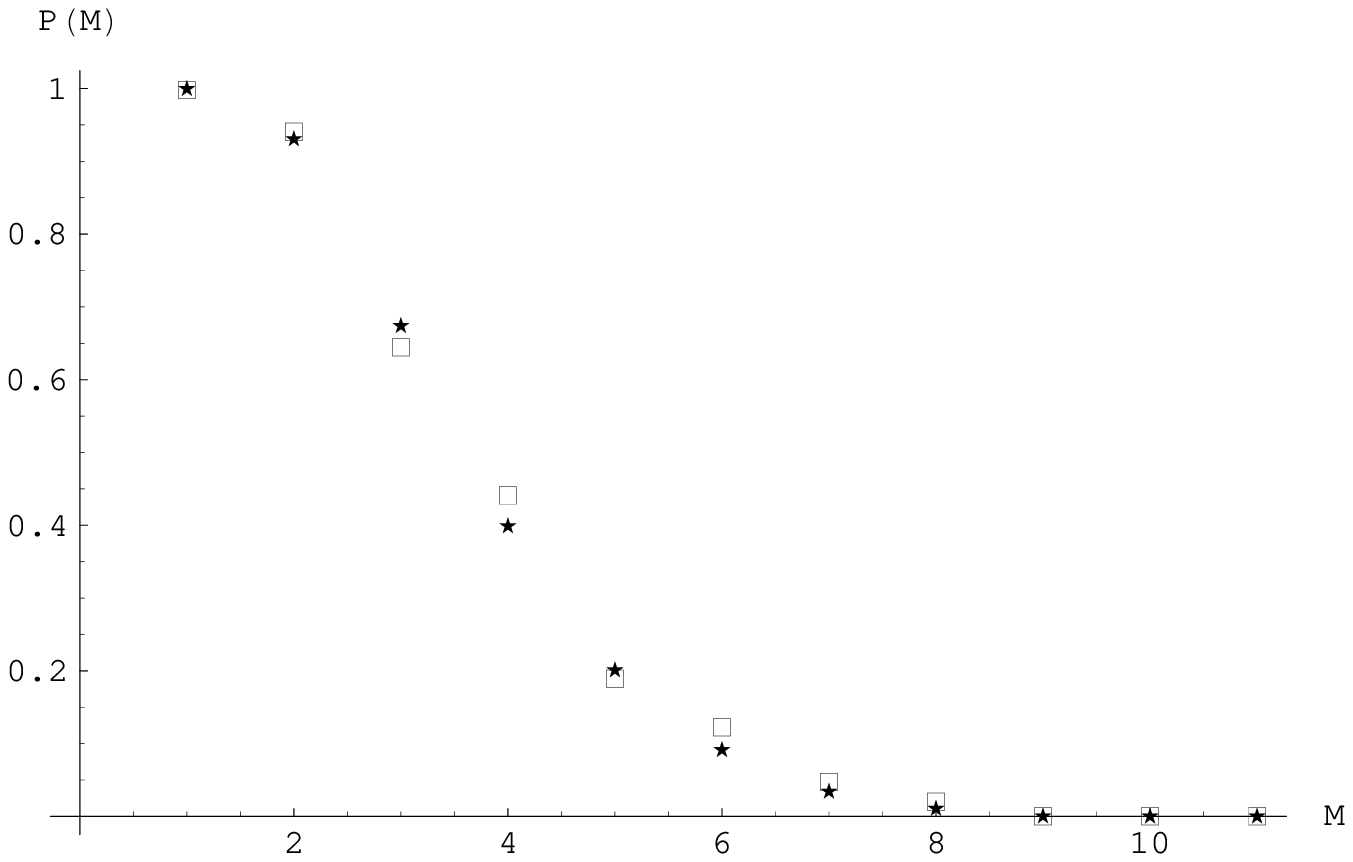}}
\caption{Probability distributions for
$L=8, U_I=1$.
The stars represent the distribution without K-theory constraints, the boxes
give the result including the constraints.}
\end{center}
\end{figure}
Using the explicit constructions obtained by the computer algorithm,
it is possible to quantify the effect of imposing the K-theory constraints
(\ref{EqKTheory}).
The overall effect of the additional constraints is to suppress the number
of possible brane configurations by a factor of five. For example in the case
without flux considered in this article, i.e. $L=8$, and for complex structures $U_I=1 \forall I$ and
non-tilted tori, we get
approximatively $1.3\times10^8$ models if we do not impose the constraints and ca.
$2.3\times10^7$ models enforcing them.

Because the distribution of models is highly dominated by configurations
where the value for $\hat{Y}^I$ is 0 or 1 (in agreement with the observation that only $1.6\%$ of all models live on tilted tori), one expects that the
constraints of equation (\ref{EqKTheory}) 
suppress models with odd total rank.
This is indeed the case as can be seen
in figure \ref{Fig_k_rkdist}.
Note that the distribution shown is limited to the case of
all complex structures chosen as  $U_I=1$, but different
values of the complex structure change only the maximum and width of the
distribution, not its shape, which is therefore generic.

By contrast, other observables are more or less unaffected by the
inclusion of K-theory constraints, for example the frequency of
models containing at least one gauge group $U(M)$, as shown in figure \ref{Fig_k_undist}
(as above the distribution for one specific value of the complex structures is
 generic).
This is due to the fact that the constraints do not limit the
individual brane configurations but only the overall composition of them to
form a consistent model. In both cases the supersymmetric brane configurations
are dominated by stacks of a small number $N_a$ of branes.

Let us compare the overall suppression of models by the K-theory constraints to the 
analogous results for Gepner Models \cite{Dijkstra:2004ym,Gato-Rivera:2005qd}. Unfortunately, a direct comparison of our gauge statistics and the Gepner Model data is not possible, since  the abundance of solutions at the Gepner point forced the authors to restrict their search a priori to Standard Model like vacua.
The effects of implementing 
also the above K-theory constraints for this class of vacua are analysed
in \cite{Gato-Rivera:2005qd}, 
resulting only in a practically negligible suppression factor.  
The discrepancy is 
even more striking if one considers that the number of symplectic branes and therefore of additional K-theory constraints
in many, though not in all of the Gepner orientifolds is much higher 
than on our toroidal orbifold. However, the MSSM-like models analysed on
the Gepner side satisfy very strong constraints, in particular the
$U(1)_Y$ massless condition (cf. equation (\ref{EqU1massless})).
We have reason to infer that these requirements a priori rule out most of those
models which violate the K-theory constraints.

\subsection{Correlations between the total rank of the gauge group and chirality}\label{sec_chivsrk}
\begin{figure}[ht!]
\begin{center}
\subfigure[with K-theory constraints]{
  \includegraphics[width=0.9\textwidth, trim=0mm 10mm 0mm 10mm, clip]{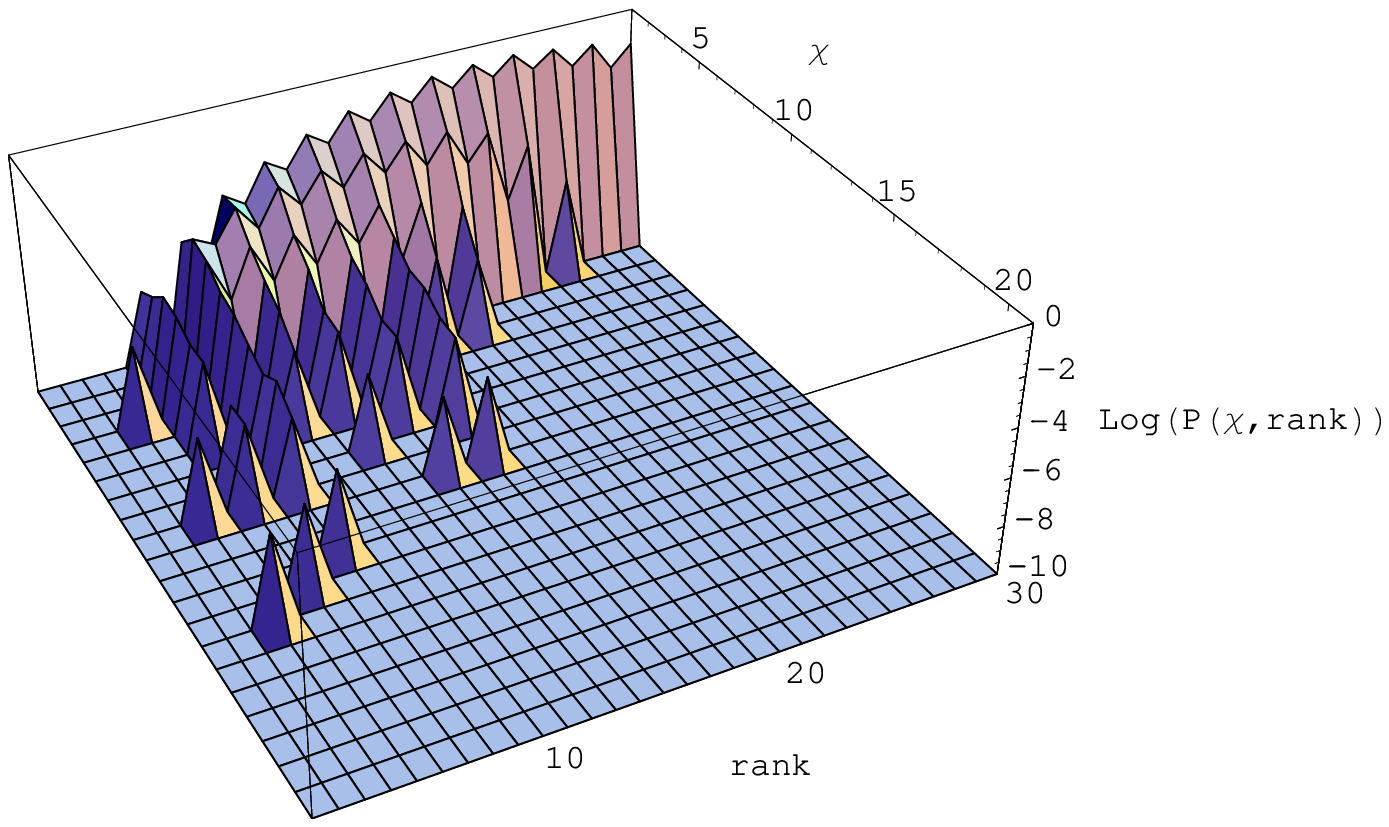}}
\subfigure[without K-theory constraints]{
  \includegraphics[width=0.9\textwidth, trim=0mm 10mm 0mm 10mm, clip]{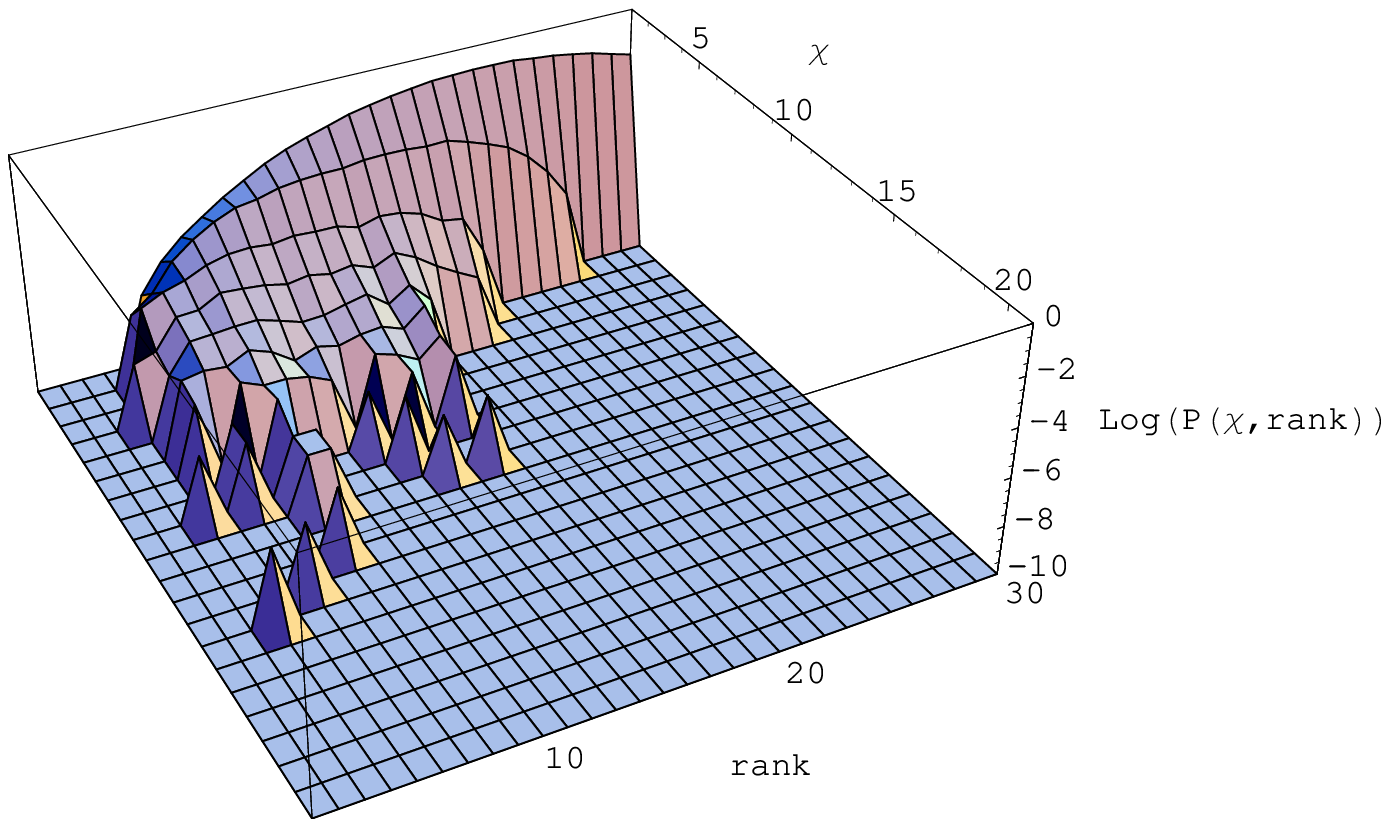}}
\caption{The frequency distribution of models of specific rank and chirality.
$L=8, U_I=1$.}
\label{Fig_k_chivsrk}
\end{center}
\end{figure}
We find a correlation
between the rank of the gauge group and the mean chirality of the model,
which we have defined as
\be\label{EqChiDef}
  \chi=<I_{a'b}-I_{ab}>=\hat{b}^{-2}<2\hat{\vec{Y}}_a\hat{\vec{X}}_b>
  =\frac{2}{k(k-1)}\sum_{\stackrel{a,b\in\{1,..,k\}}{b<a}}\hat{b}^{-2}\hat{\vec{Y}}_a\hat{\vec{X}}_b.
\ee
In this respect the explicit computer search confirms the
results obtained via a saddle point approximation in \cite{Blumenhagen:2004xx}.

As can be seen in figure \ref{Fig_k_chivsrk}, the K-theory constraints do not change the overall shape of the distribution except for a suppression of models with odd total rank, as already seen in figure \ref{Fig_k_rkdist}.

\subsection{Statistics of Standard-like models}\label{sec_mssmstat}

\begin{figure}[hbt]
\begin{center}
\subfigure[MSSM models]{\label{Fig_gensm}\includegraphics[width=0.9\textwidth]{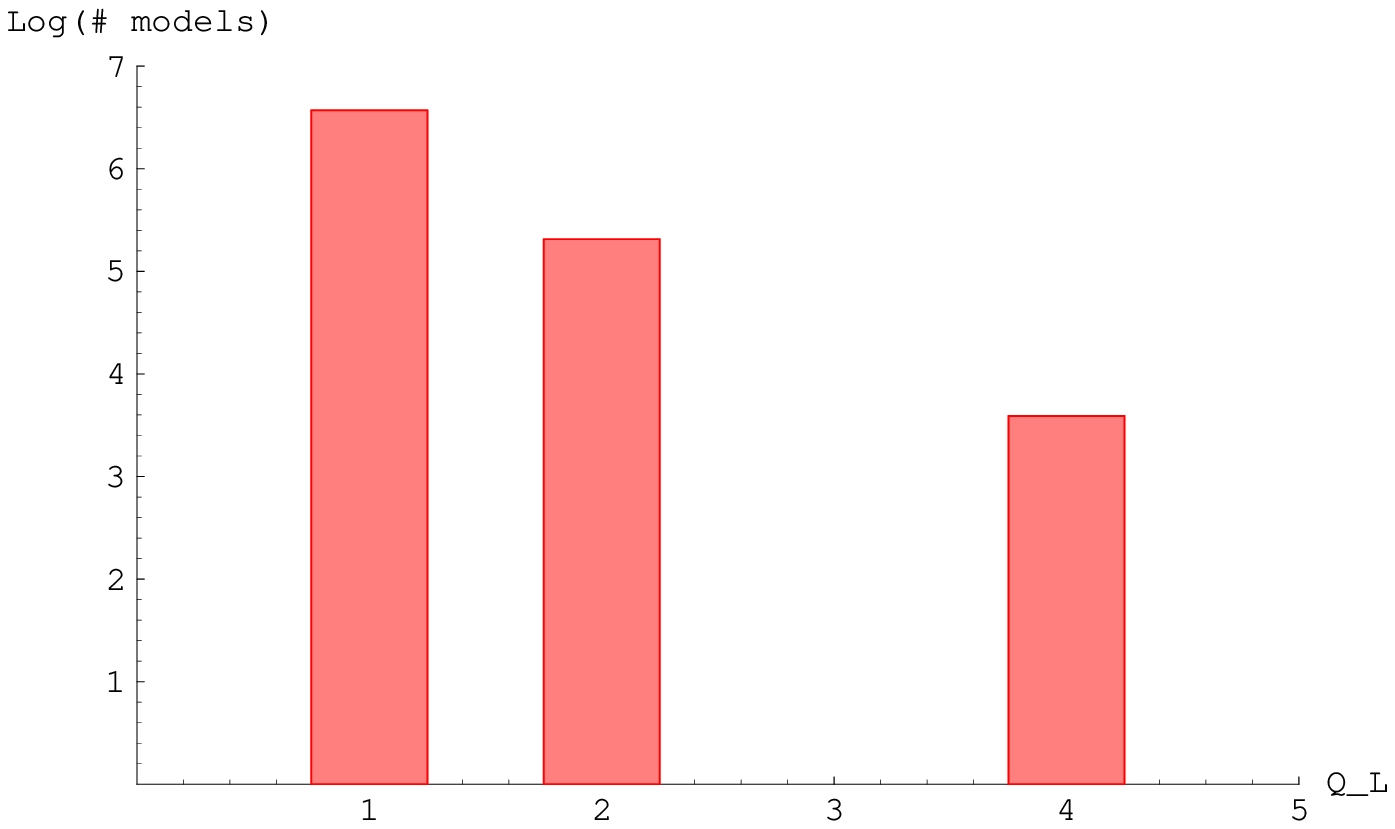}}
\subfigure[Pati-Salam models]{\label{Fig_genPS}\includegraphics[width=0.9\textwidth]{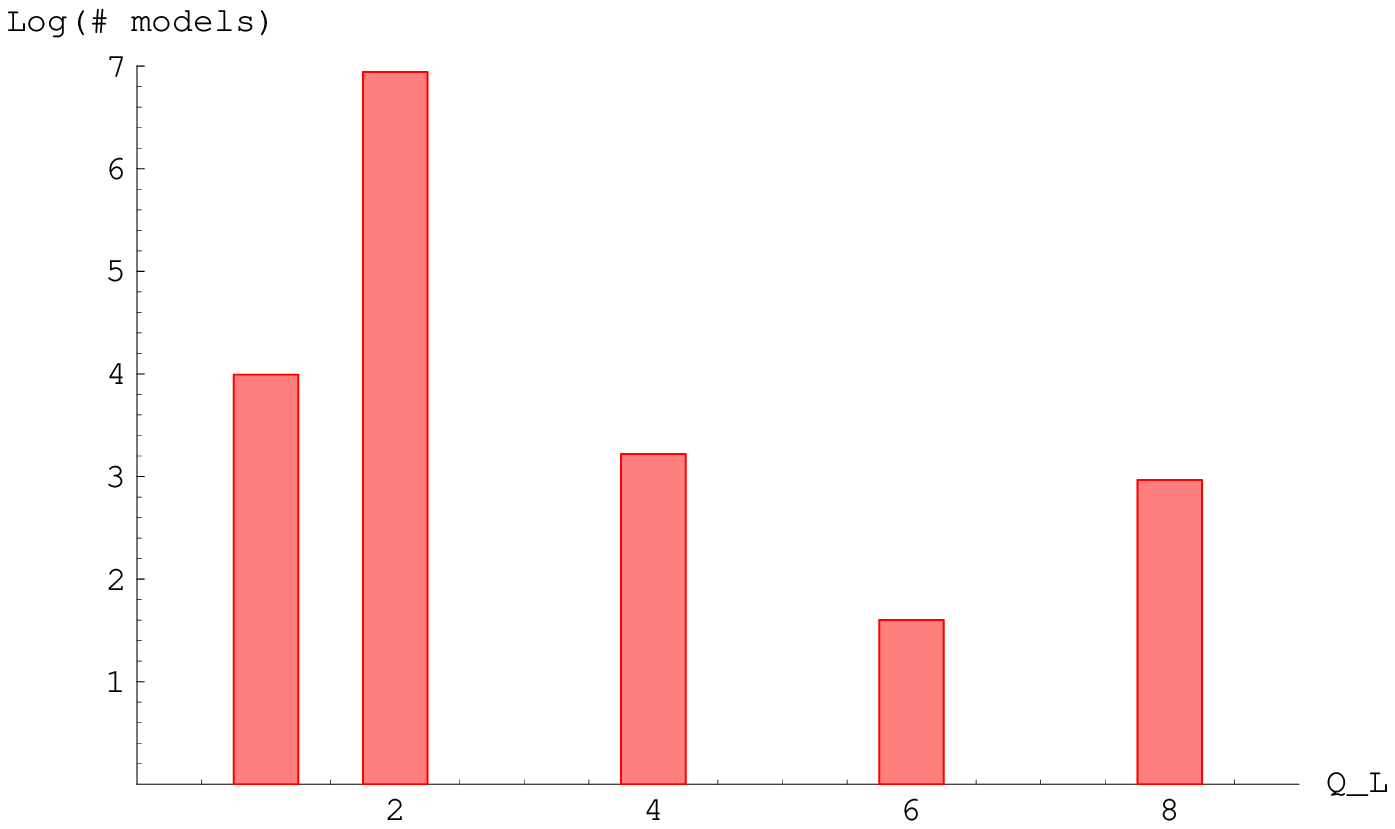}}
\caption{Logarithmic plot of the number of models
  versus the number of generations.}
\end{center}
\end{figure}

An analysis of the possible realizations of models with MSSM-like features, as described in section \ref{subsec_smtypes}, shows a 
surprising result: We do not encounter any three-generation Standard Model in our analysis.
As displayed in figure \ref{Fig_gensm} the only configurations we have found exhibit one, two
or four generations, with a strong statistical domination of the
one-generation models.

Given the fact that models with three generations have been constructed
explicitly in our setup (see e.g.\cite{Marchesano:2004xz, Cvetic:2005bn}) this might appear as an incorrect result at first sight.
The important point to notice here is that all three-generation models
known to us share the property that they need values for the complex
structures which in our conventions are very large and out of the reach
of our computer analysis.
Another issue not to be neglected is that most of the models
considered
in the literature make use of some special features
like brane recombination,
brane splitting (breaking of higher rank gauge groups) etc.
From this observation and the arguments in
section \ref{sec_numsol} we might draw the conclusion that three-generation
models are statistically very highly suppressed in this specific orbifold
setup.

\subsection{Massless hypercharge}
\begin{figure}[htb!]
\hspace{-5mm}\includegraphics[width=1.1\textwidth, trim=0mm 10mm 0mm 10mm, clip]{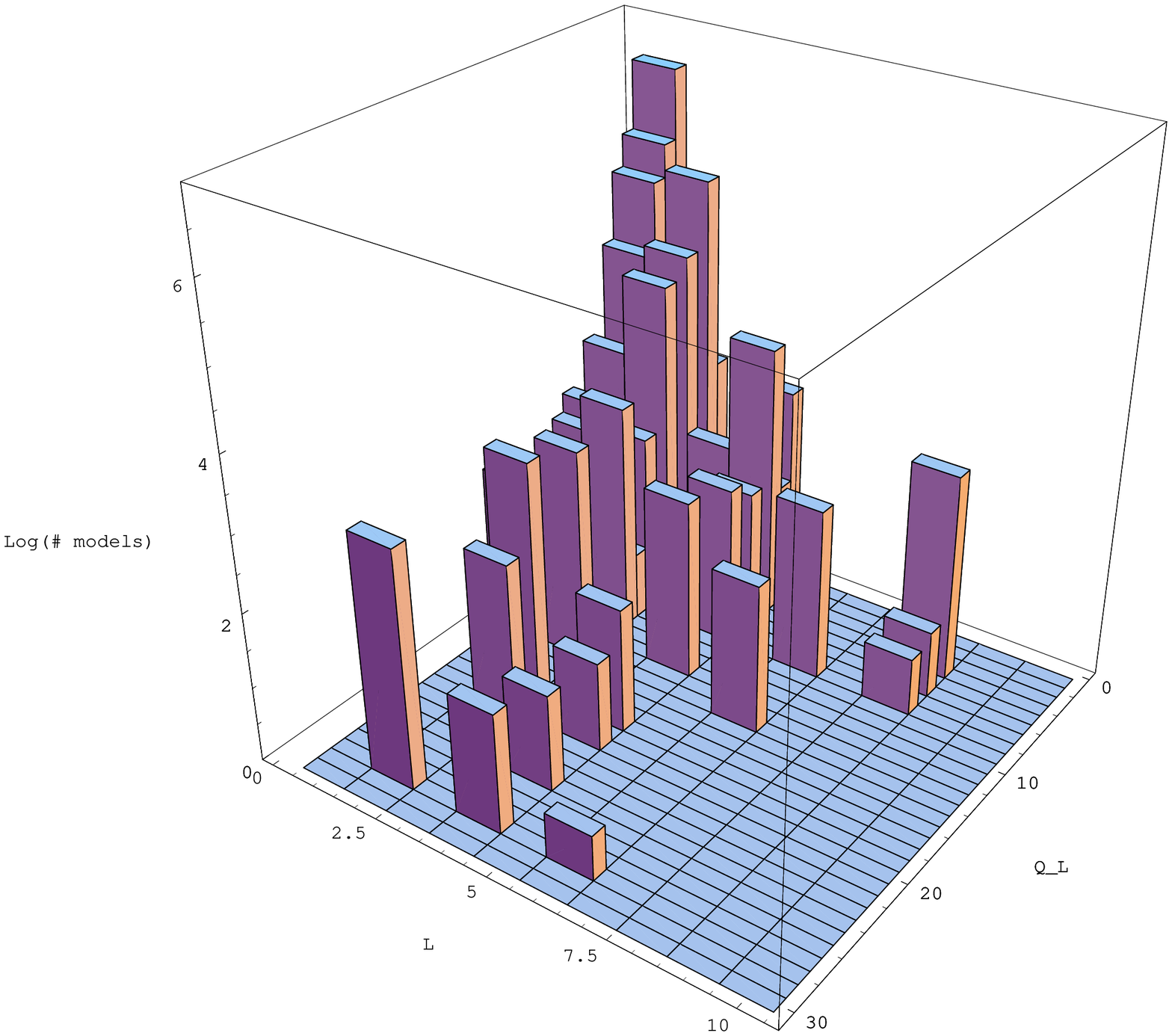}
\caption{Logarithmic plot of the number of models with Standard Model
  characteristics, not including the condition for a massless hypercharge,
  depending on the number of generations in the quark ($Q_L$)
  and lepton ($L$) sector.}
\label{Fig_genmu1}
\end{figure}
To gain some further insight into the distribution of Standard Model-like
properties, we have performed an analysis of our models that includes all constraints
for MSSM-like models except for  the condition of having a massless hypercharge $U(1)_Y$.
Furthermore we allow for a different number of generations in the quark
and lepton sector. The resulting distribution for different numbers of
quark and lepton generations in shown in figure \ref{Fig_genmu1}.

From the phenomenological point of view these models are of course not
extremely useful because the massless $U(1)_Y$ condition is important for
a consistent MSSM structure, but the result is nevertheless quite
interesting, as it turns out that even within this  relaxed framework  no
three generation model can be found. The closest we can get to a three-generation model
is a configuration with three generations of quarks and four lepton
generations.

\subsection{Pati-Salam models}
In addition to the direct realisation of MSSM models we also analysed
the occurrence of models with a Pati-Salam gauge group
\be
  SU(4) \times SU(2)_L \times SU(2)_R.
\ee
As can be seen in figure \ref{Fig_genPS}, we get one-, two-, four-, six-
and eight-generation models, but again no model with three generations.
Also in this case there exist explicit constructions of three-generation
Standard Models in the literature
\cite{Cvetic:2005bn}
and our conclusions are the same as in section \ref{sec_mssmstat} - the
statistical suppression of three generation models is extremely large.

\subsection{Statistics of the hidden sector}\label{sec_hidden}
\begin{figure}[ht!]
\begin{center}
\subfigure[Rank distribution]{
  \label{Fig_hiddenrk}\includegraphics[width=0.9\textwidth]{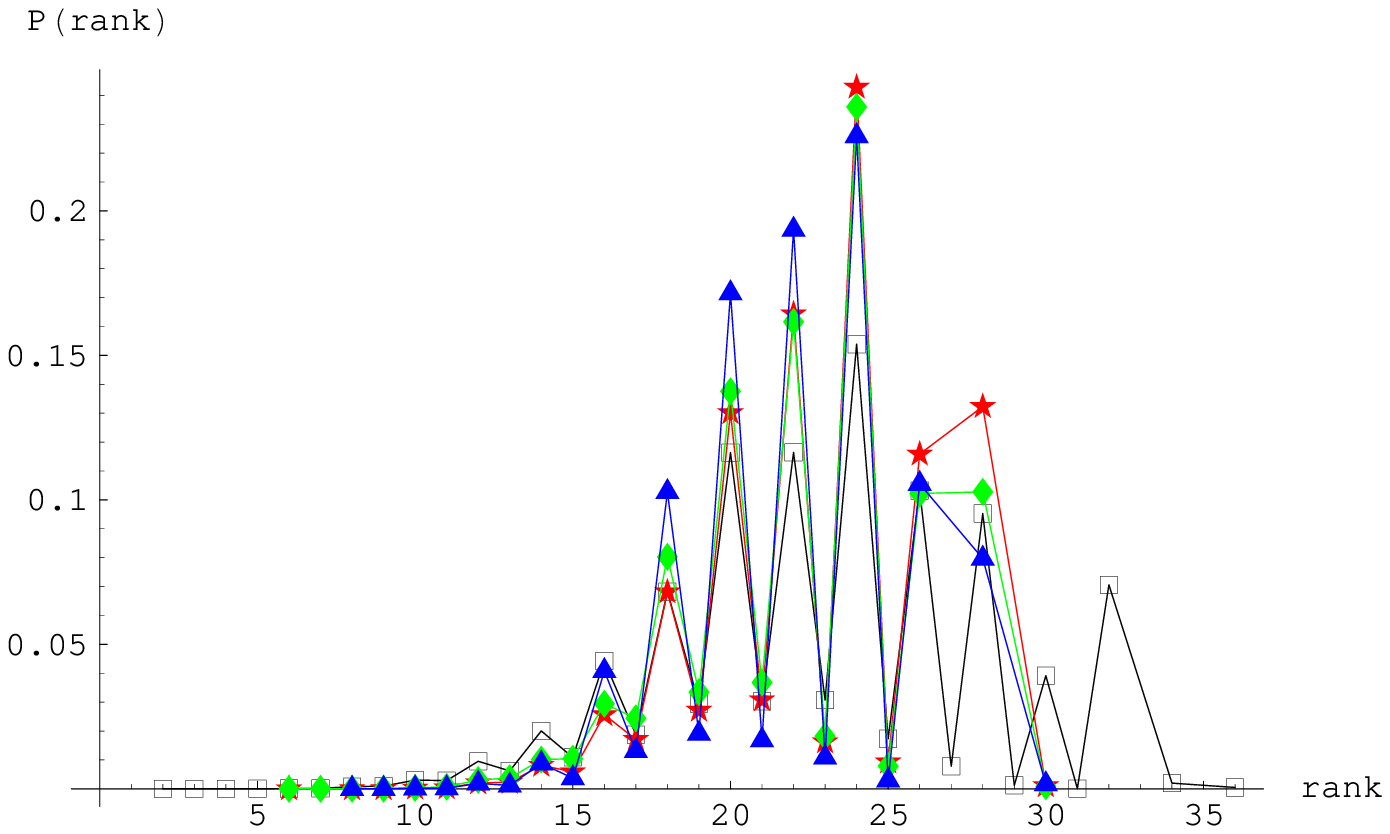}
}
\subfigure[$U(M)$ distribution]{
  \label{Fig_hiddenun}\includegraphics[width=0.9\textwidth]{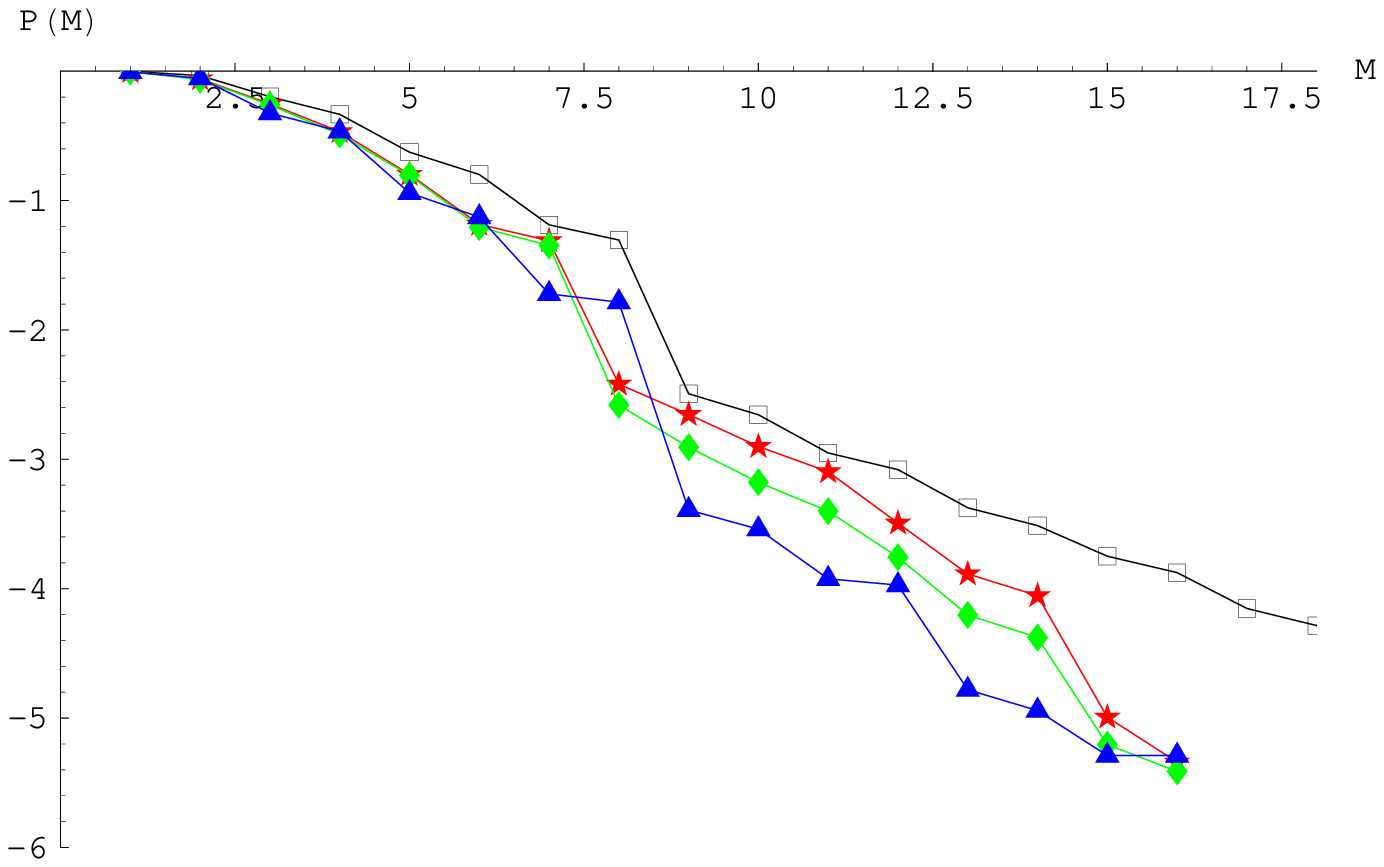}
}
\caption{
Frequency distributions of all models (empty black boxes) and in the
hidden sector of
MSSM models (red stars), MSSM models without restriction to a massless
$U(1)$ (green diamonds)
and Pati-Salam models (blue triangles).
}
\end{center}
\end{figure}
An interesting question in a statistical context concerns the properties of the
hidden sector of our models.
In figures \ref{Fig_hiddenrk} and \ref{Fig_hiddenun} the distribution of the
total rank of the gauge group and the frequency of models containing a $U(M)$
gauge
factor in the hidden sector are compared with the results one obtains from
an analysis of all models considered.

As it turns out the distributions
of the constrained models differ only very little from those of the full
set.
The small deviations which are visible in the plots cannot be regarded as indications of
fundamentally different behaviour. They must be seen rather as a remnant of
the statistical analysis, which looses its applicability in the region
of very high rank of the gauge groups where the number of models is very small as compared
 to the region of lower rank.

This suggests that the statistical distributions are not affected
by the additional constraints of requiring a specific configuration in the
visible sector.
This observation gives some hints about the correlation of variables, which will
be explored in greater detail in the section \ref{sec_corr}.

\begin{figure}
\begin{center}
\subfigure[Full set of models]{
  \label{fig_gdim1}\includegraphics[width=0.9\textwidth]{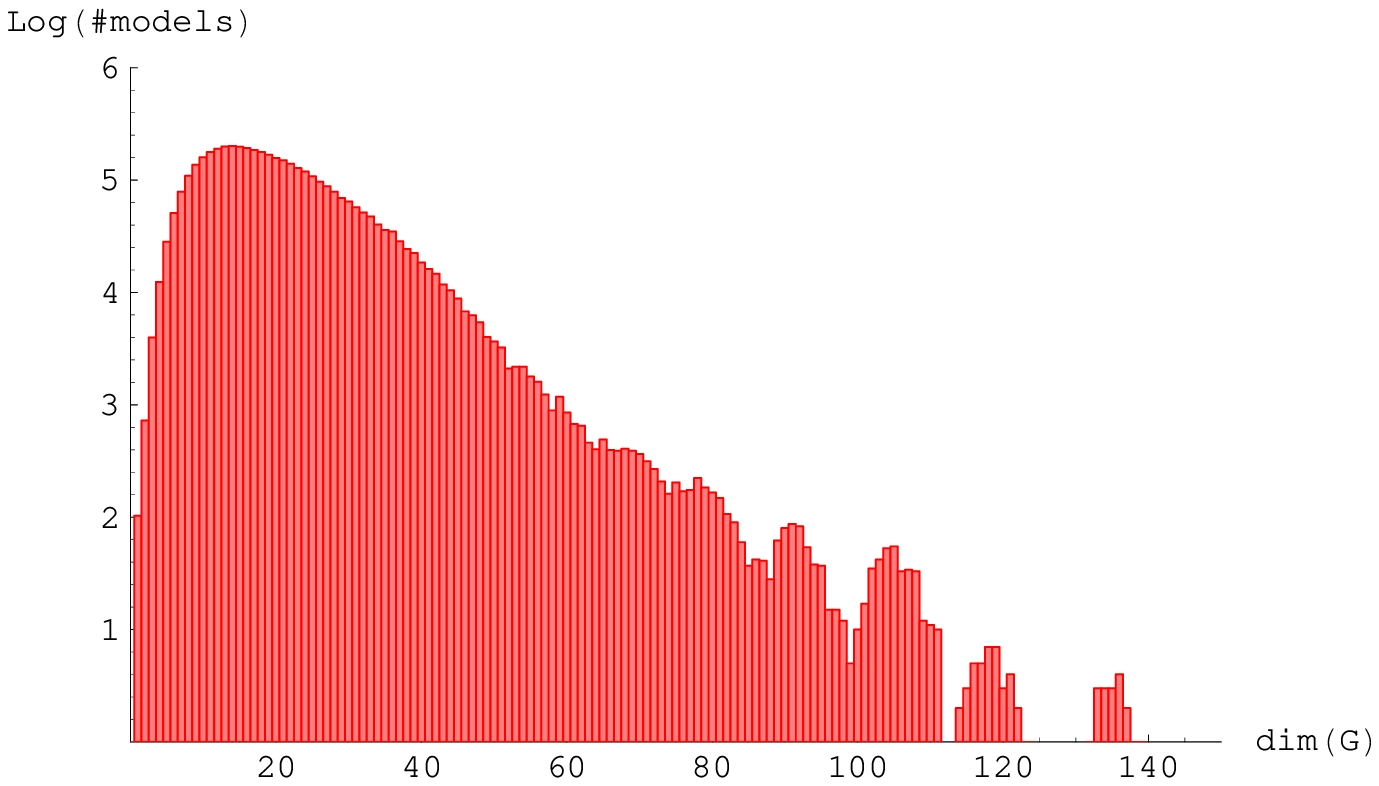}
}
\subfigure[Restricted set with max. three branes in the hidden sector]{
  \label{fig_gdim2}\includegraphics[width=0.9\textwidth]{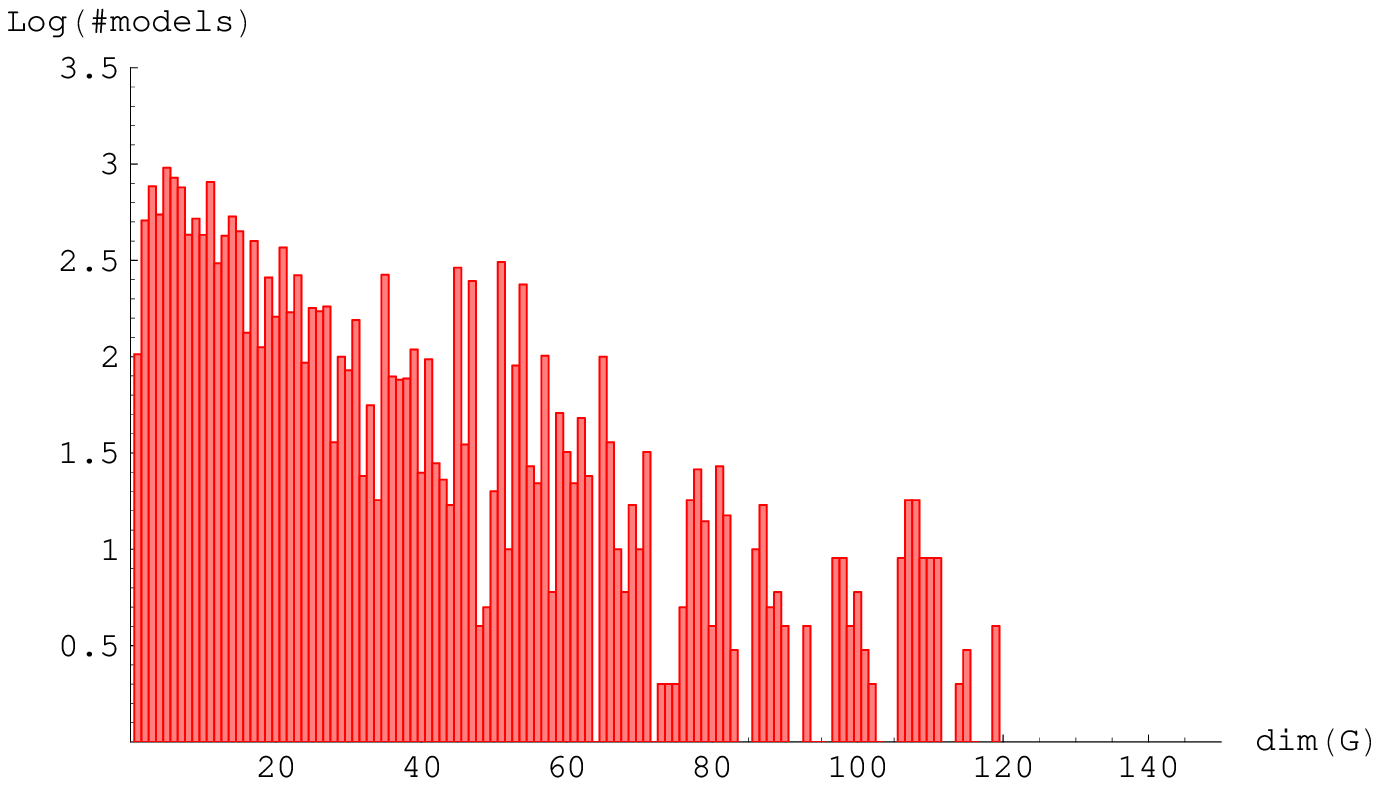}
}
\caption{
Logarithmic plots of the number of models with Standard Model gauge group and
massless $U(1)$ depending on the dimension of the gauge group G in the hidden
sector.
}
\end{center}
\end{figure}
A comparison of the distribution of the total dimension of the gauge group in
the hidden sector of Standard Model-like configurations with the results for
Gepner models in \cite{Dijkstra:2004cc} shows two quite similar pictures.
In figure \ref{fig_gdim1} we give the complete distribution, in \ref{fig_gdim2}
we allow for maximally three branes in the hidden sector in order to make
the result better comparable to \cite{Dijkstra:2004cc}, where a similar
cutoff was imposed for computational reasons.
The main difference comes from the fact that we have much fewer models in our
ensemble and, more importantly, our models are not constrained to three
generations.

\subsection{Gauge couplings}\label{sec_gcoup}
\begin{figure}[htb]
\begin{center}
\includegraphics[width=0.9\textwidth]{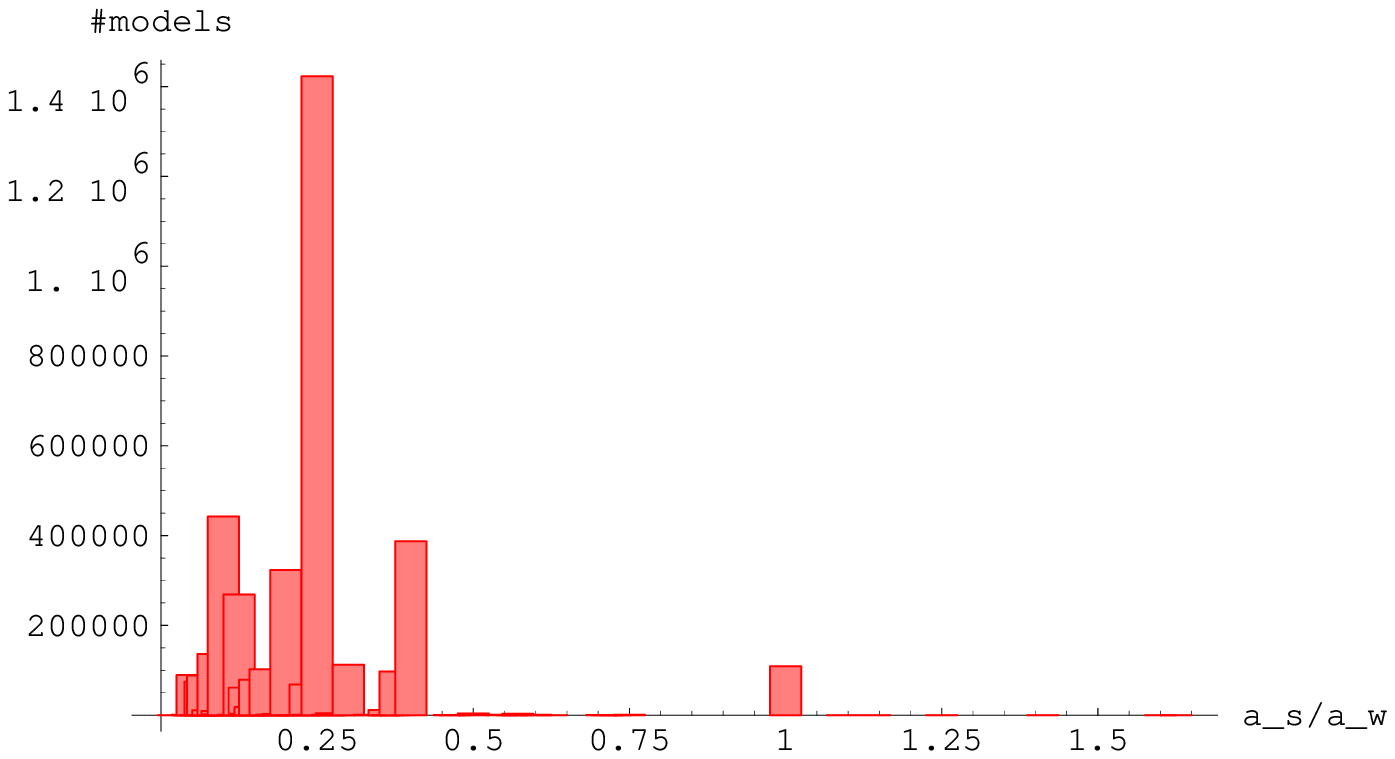}
\caption{Distribution of $\frac{\alpha_s}{\alpha_w}$ in Standard Model-like
configurations.}
\label{fig_gcdist}
\end{center}
\begin{center}
\includegraphics[width=0.9\textwidth]{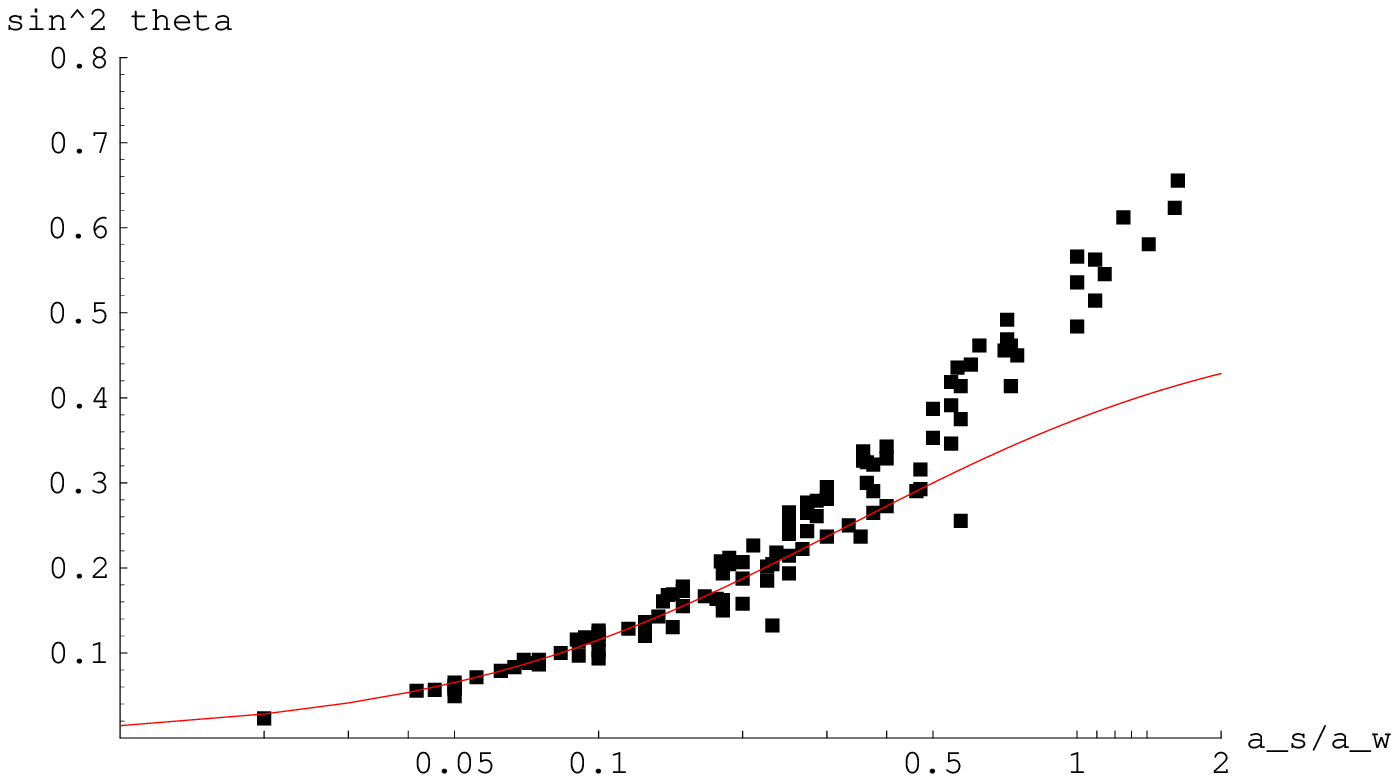}
\caption{Values of $sin^2\theta$ depending on $\frac{\alpha_s}{\alpha_w}$
(on a logarithmic scale). Each dot represents a class of models with these
values. The red line shows the relation (\ref{eq_ccrel}), which is satisfied for $88\%$ of all models}
\label{fig_sint}
\end{center}
\end{figure}

The observables considered so far all belong only to the topological sector of the model in that they are defined by the wrapping numbers of the branes and manifestly independent of the geometric moduli of the compactification manifold. A quantity which is relatively easy to analyse but which does contain geometrical information  are the gauge couplings in the visible sector.
In \cite{Blumenhagen:2003jy} it was argued that a MSSM model built with
intersecting branes naturally leads to a relation between the coupling
constants,
\be\label{eq_ccrel}
  \frac{1}{\alpha_Y}=\frac{2}{3}\frac{1}{\alpha_s}+\frac{1}{\alpha_w},
\ee
where $\alpha_{Y/s/w}$ are the couplings of the hypercharge, the strong and
weak sector respectively.
Note that this relation of the coupling constants does not imply that these models automatically exhibit
full gauge unification,
but  it could fit into an $SU(5)$
framework.

For an honest check of this relation we would have to use the
renormalization group equations and evolve the coupling constants from the
string scale down to their low energy values.
In our analysis we have not done so, but investigated instead the relation between
the couplings at the string scale. Therefore we can only get some hints
towards the actual relation at lower scales.

To calculate the gauge couplings we use the expression derived in
\cite{Blumenhagen:2003jy} for the gauge coupling of a stack $a$ in an
intersecting brane model,
which reads in our conventions
\be\label{gceq}
  \frac{1}{\alpha_a}=\frac{c}{\kappa_a}\,\frac{1}{\hat{b}\sqrt{\prod_{i=1}^3R_1^{(i)}R_2^{(i)}}}\sum_{I=0}^{3}\hat{X}^I U_I.
\ee
The constant $c$ is given by
\be
  c=\frac{1}{2\sqrt{2}}\frac{M_{Planck}}{M_s}
\ee
and $\kappa_a$ is 1 for an $U(N)$ stack and 2 for a $SO(2N)$ or $Sp(2N)$ stack. Note the explicit dependence on the complex structures.

For the coupling of the hypercharge $\alpha_Y$, we have to consider the
contribution from all stacks of branes in the visible sector and distinguish
the three different possibilities we consider for a MSSM configuration, as
explained in section \ref{subsec_smtypes}. Using
\be
  \frac{1}{\alpha_Y}=\sum_i 2N_ix_i^2\frac{1}{\alpha_i},
\ee
we get for the three configurations
\be
  \frac{1}{\alpha_Y}=
  \left\{\begin{array}{rl}
    1. & \displaystyle
         \frac{1}{6}\frac{1}{\alpha_a}+\frac{1}{2}\frac{1}{\alpha_c}+
         \frac{1}{2}\frac{1}{\alpha_d},\\[2ex]
    2. & \displaystyle
         \frac{2}{3}\frac{1}{\alpha_a}+\frac{1}{\alpha_b},\\[2ex]
    3. & \displaystyle
         \frac{2}{3}\frac{1}{\alpha_a}+\frac{1}{\alpha_b}+
         2\frac{1}{\alpha_d}.
  \end{array}\right.
\ee

In figure \ref{fig_gcdist} the distribution of the ratio $\alpha_s$ to $\alpha_w$
is displayed.
We find that only for 2.75\% of all models  $\alpha_s = \alpha_w$ at the string scale and that the weak coupling
constant is generically larger than the strong one.

Figure \ref{fig_sint} shows the different values for $\sin^2\theta$,
depending on
the ratio of $\alpha_s/\alpha_w$. The red line represents the ratio given in
(\ref{eq_ccrel}). $\sin^2\theta$ is calculated from
\be
  \sin^2\theta=\frac{\alpha_Y}{\alpha_Y+\alpha_w}.
\ee
We find that 88\% of the models obey the relation (\ref{eq_ccrel}).
Note that this result might be a bit obscured by the plot
because it shows one dot for every possible value, not taking into account
that there exist many models with exactly the same values.
In fact there exist
many more models with low values for $\alpha_s/\alpha_w$ and  $\sin^2\theta$,
which happen to be the ones fulfilling the relation.

These results can also be compared with the analysis
of \cite{Dijkstra:2004cc}, where as in the case of the hidden gauge group
one has to take into account that we are dealing with a smaller ensemble
and with models that are not constrained to exactly three generations of
fermions. The fraction of MSSM-like Gepner models satisfying (\ref{eq_ccrel}) is found there to be only ca. $10\%$. Without wanting to over-interpret this mismatch, we emphasize once more  that the gauge couplings are not part of the topological sector of the theory and that therefore deviations between the small and large radius regime fit into the general picture.

\section{Correlations}\label{sec_corr}

\begin{figure}[ht]
\begin{center}
\subfigure[MSSM-like models]{
  \includegraphics[width=0.5\textwidth, trim=0mm 10mm 0mm 10mm, clip]{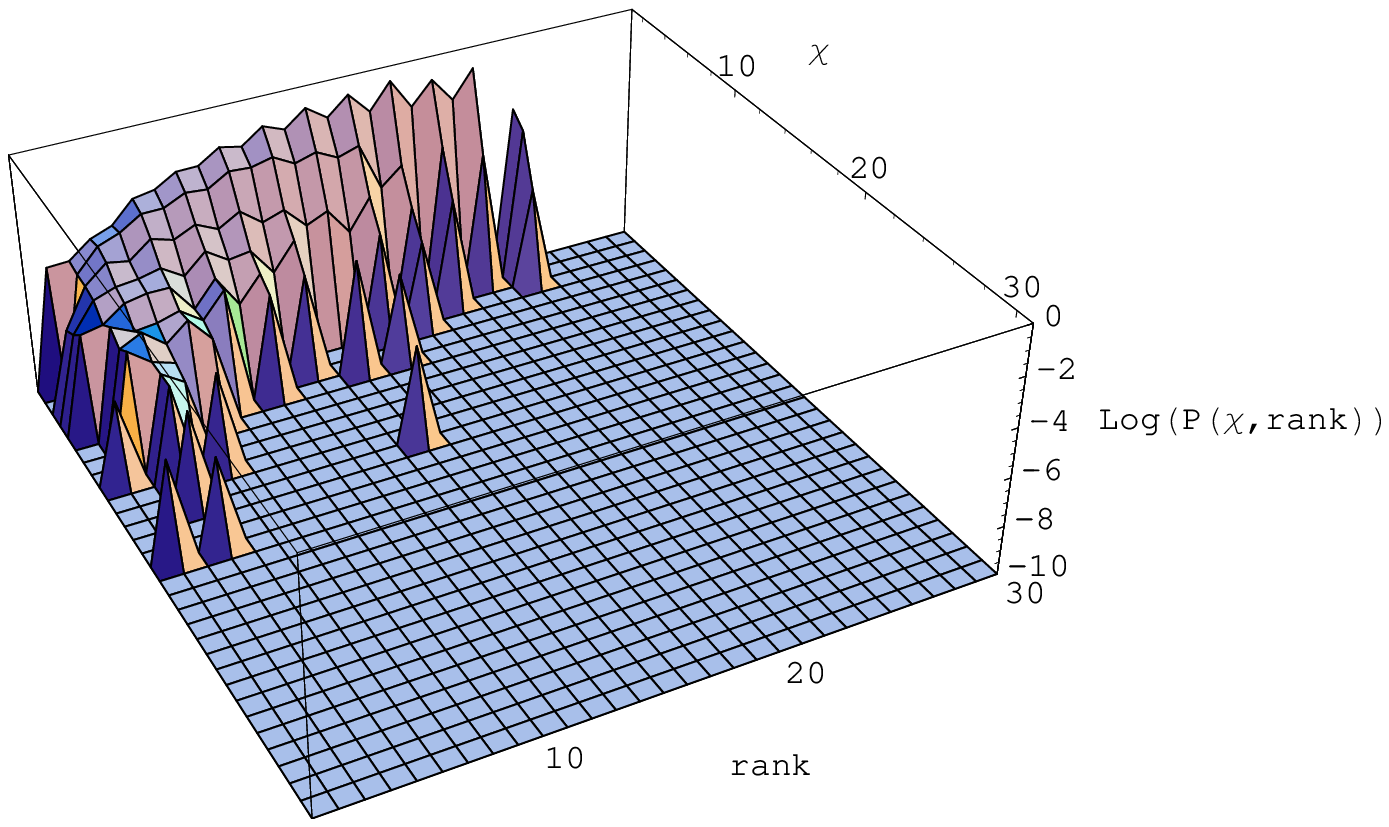}
}
\subfigure[MSSM-like models with massive $U(1)$]{
  \includegraphics[width=0.5\textwidth, trim=0mm 10mm 0mm 10mm, clip]{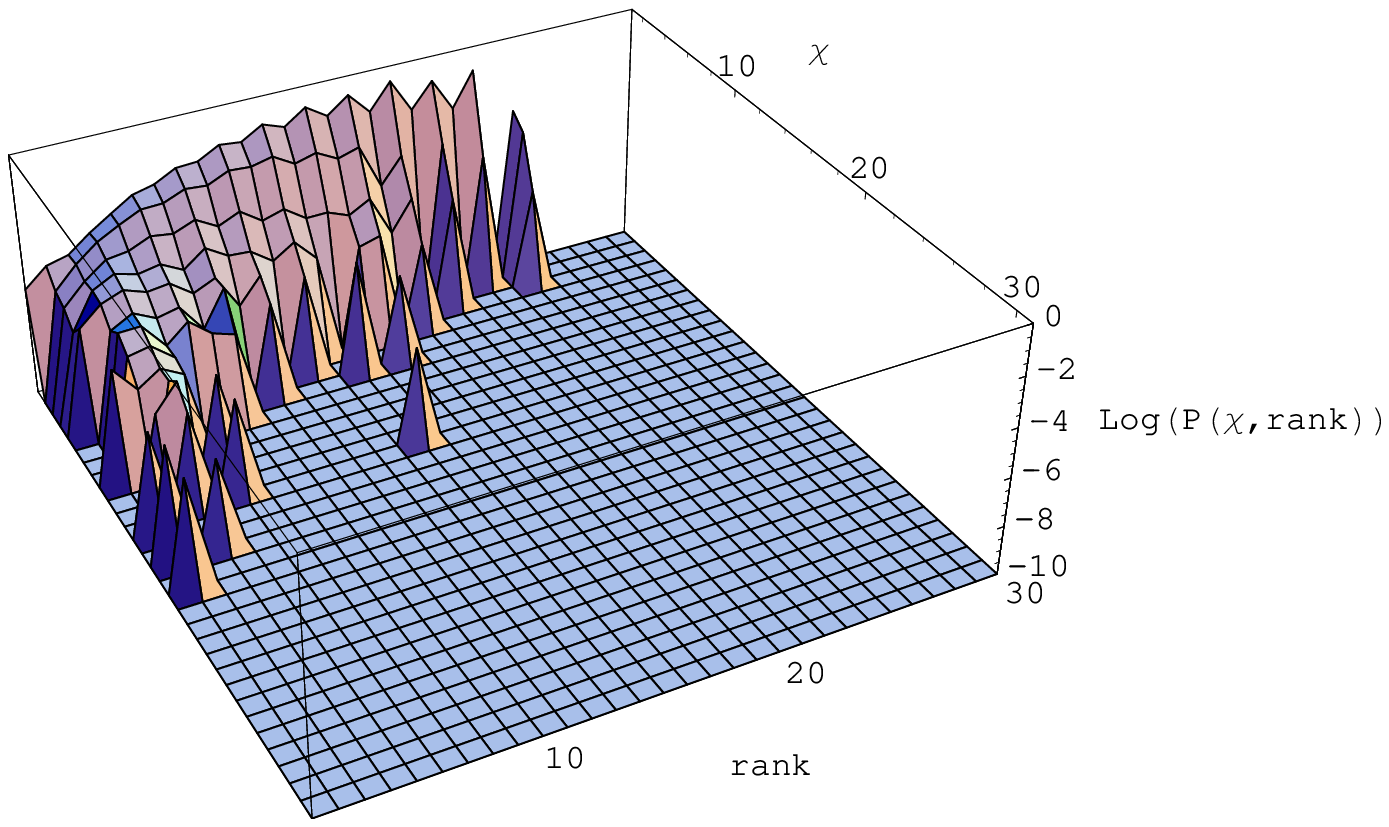}
}
\subfigure[Pati-Salam models]{
  \includegraphics[width=0.5\textwidth, trim=0mm 10mm 0mm 10mm, clip]{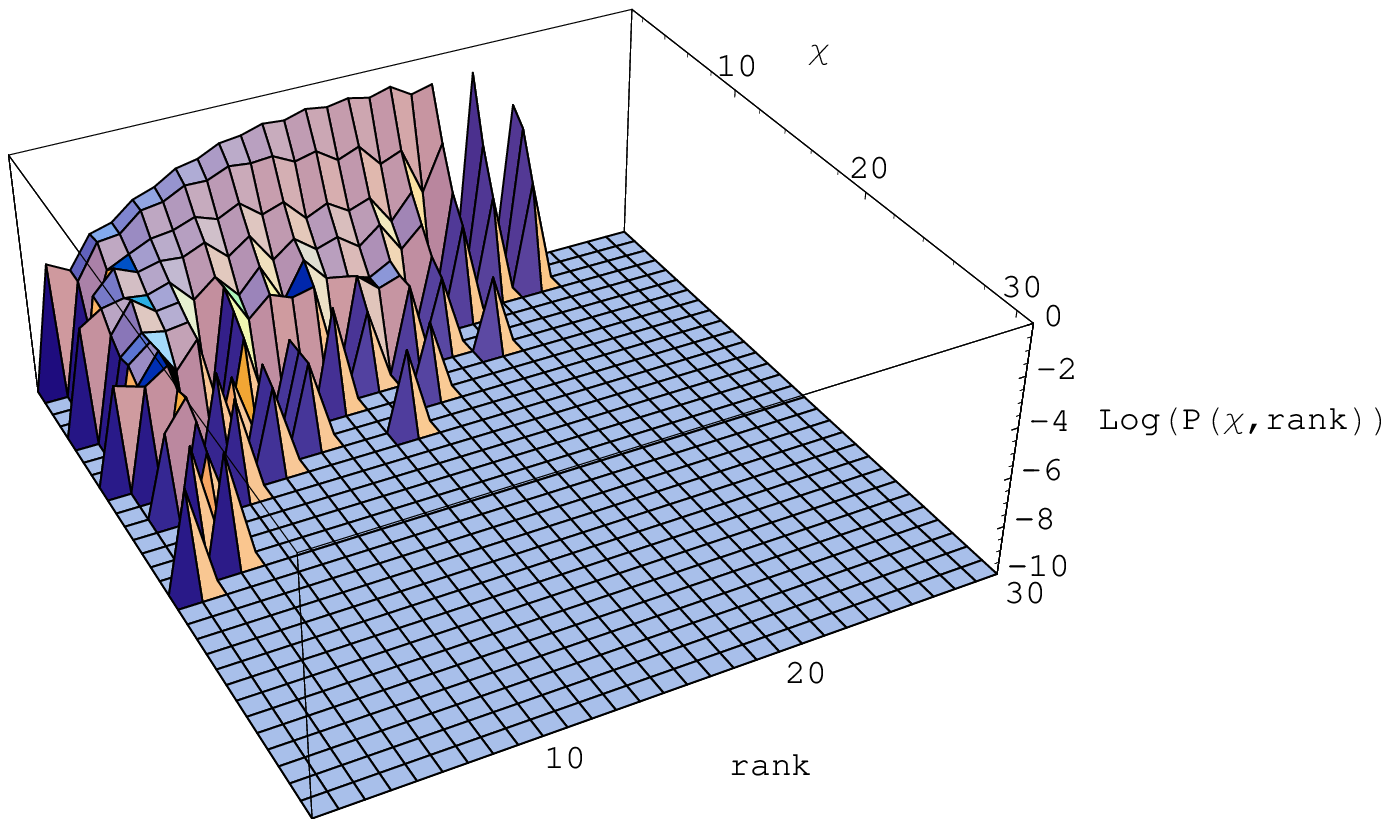}
}
\caption{The relative frequency to find models of specific rank and chirality,
  requiring different visible sectors to be present in the models.}
\label{fig_chivsrkall}
\end{center}
\end{figure}
\begin{figure}[ht]
\begin{center}
\subfigure[$\scriptstyle P_c=|P(3)*P(2/Sp2)-P(2/Sp2\wedge3)|/C$]{
  \label{Fig_U23cor}\includegraphics[width=0.9\textwidth]{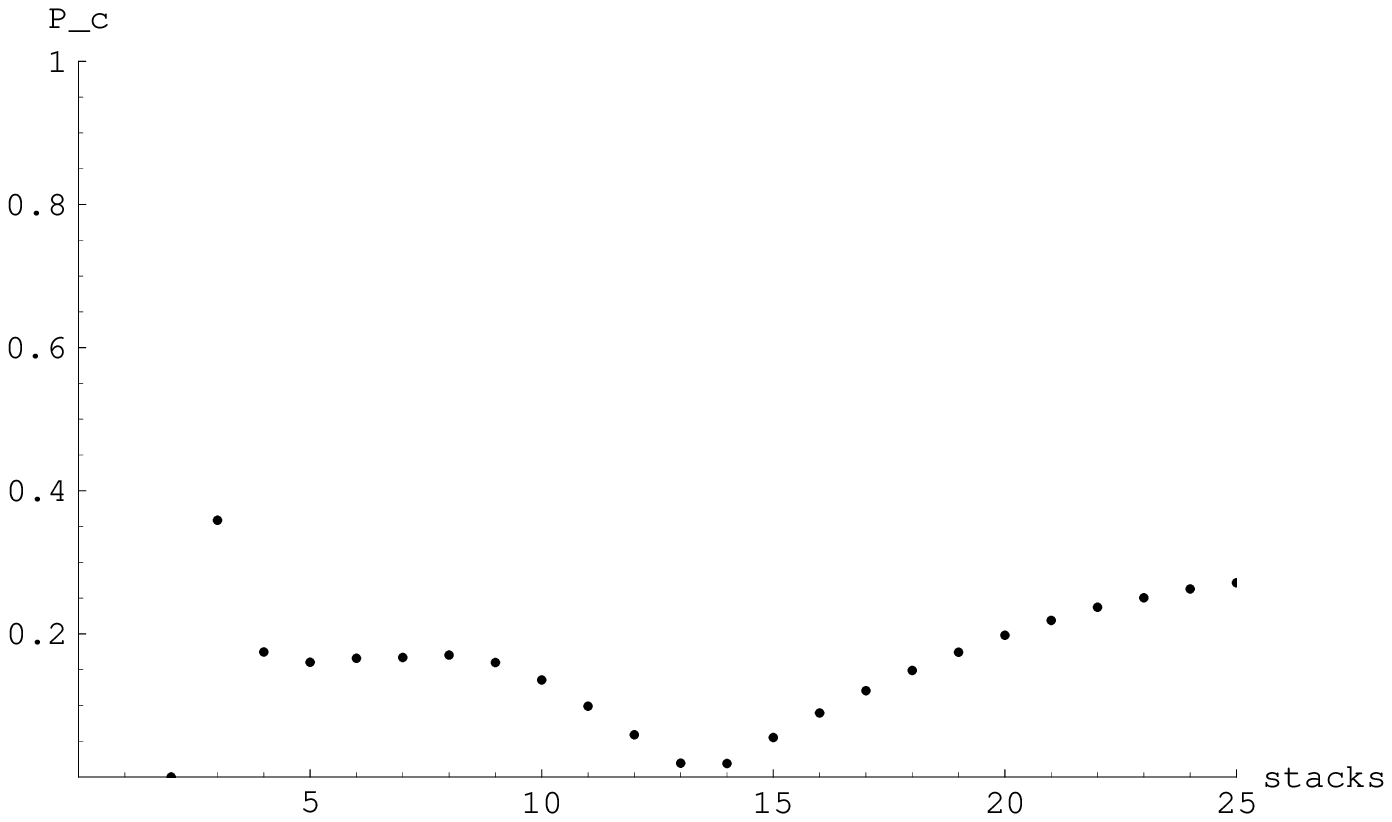}
}
\subfigure[$\scriptstyle P_c=|P(3)*P(ns)-P(3_{ns})|/C$]{
  \label{Fig_U3symcor}\includegraphics[width=0.9\textwidth]{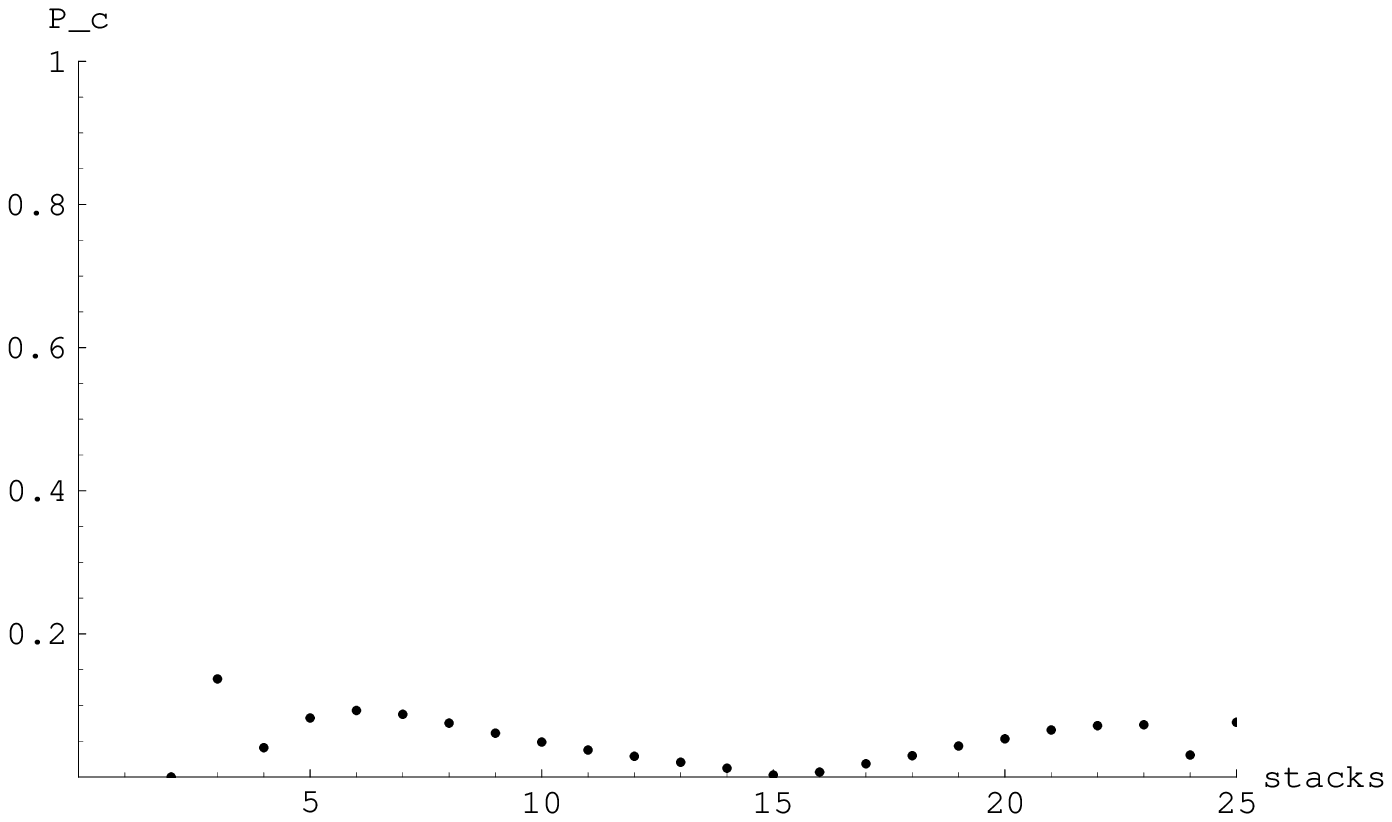}
}
\caption{
Correlations between atomic properties of MSSM-like configurations.
The factor $C$ is a normalization, see equation (\ref{eq_pc}).
Relative frequencies are labeled as follows:
$P(M)$: frequency of stacks with gauge group $U(M)$;
$P(2/Sp2)$: frequency of $U(2)$ or $Sp(2)$ stacks;
$P(ns)$: frequency of gauge groups without symmetric representations.
}
\label{Fig_corr}
\end{center}
\end{figure}

In this section we investigate the
correlation between different observables in greater detail.
As analysed
in section \ref{sec_hidden}, the statistics of the hidden sector of MSSM-like
and Pati-Salam models are, on a statistical basis, very similar to the full
statistics of all models.
The same is also true for the
non-trivial correlations between
observables in our models.
This is shown in figure \ref{fig_chivsrkall} for the example of the correlation
between mean chirality and 
total rank of the gauge group, as discussed in section
\ref{sec_chivsrk}.

Putting these hints together we might conclude that the `atomic'
frequency distributions actually do not depend on the specific kind 
of models chosen (e.g. MSSM-like, Pati-Salam or something else), but only
on a generic distribution which is already present in the completely
unconstrained setup.

To make this statement more quantitative, we calculate the
frequencies for some atomic properties important for model building.
In particular, we check whether two properties $A$ and $B$
are statistically independent in the ensemble of D-brane models. A good measure for statistical 
dependence is given by
\be\label{eq_pc}
     P_{AB}=\frac{P(A)\times P(B)-P(A\wedge B)}{P(A)\times P(B)+P(A\wedge B)}.
\ee

As properties we choose the appearance of models with at least one
$U(3)$ or $U(2)/Sp(2)$ gauge factors
or with  at least one stack without symmetric representations.

The results are presented in figure \ref{Fig_corr} depending on the number
of stacks of the models under consideration. Note that 
the frequency of finding $U(1)$ stacks is close to one and therefore
does not lead to any further suppression for the appearance of a MSSM like
model. The plots demonstrate that, for sufficiently large 
numbers of stacks, the considered properties are statistically
independent and that therefore their frequencies simply multiply.
Note that this was an assumption in the statistical estimates
carried out in the original work \cite{Douglas:2003um}.
The combined values (integrated over all models)
for the correlation between the properties of an
MSSM-like model (without restriction to massless $U(1)_Y$) is 0.094, which is
reasonably low to justify considering the individual properties
as essentially independent. 

\subsection{Estimate of three generation Standard Models}
Having found that the constraints are statistically only very little
correlated we can make some predictions about
the overall probability of configurations in our setup which do not show
up in the data like for example true Standard Models.
In table \ref{tab_corr} we summarize the factors coming from the various
constraints. The two $U(1)$ gauge groups required for our Standard Model setup
are not included, because, as mentioned, the frequency
of having one of them is essentially one. In addition we give the fraction of models - among all configurations containing a $U(3) \times U(2)$ factor without symmetric representations - 
which meet in addition the individual constraints of a massless hypercharge, of three generations of quarks or of three  lepton families. 

\begin{table}[htb!]
\begin{center}
\begin{tabular}{|l|r|}\hline
Restriction                    & Factor\\\hline
gauge factor $U(3)$            & $0.0816$\\
gauge factor $U(2)/Sp(2)$      & $0.992$\\
No symmetric representations   & $0.839$\\
Massless $U(1)_Y$	       & $0.423$\\
Three generations of quarks    & $2.92\times10^{-5}$\\
Three generations of leptons   & $1.62\times10^{-3}$\\\hline
\emph{Total}                   & $1.3\times10^{-9}$\\\hline
\end{tabular}
\caption{Suppression factors for various constraints of Standard Model
properties.}
\label{tab_corr}
\end{center}
\end{table}
Multiplying  all these factors leads to an overall suppression factor of
\bea
  R\approx\,1.3\times10^{-9},
\eea
i.e. only {\it one in a billion} models gives rise to a four stack D-brane vacuum
with  Standard Model gauge symmetry and three generations of quark and leptons. 
Multiplying this with the total number of models
analysed,
$N\approx\,1.66\times10^{8}$, leaves us with 0.21 true Standard Models in
our ensemble.

\subsection{How good is this estimate?}
To get some idea about the quality of this estimate we can apply the
method used for three generation models to Standard Model-like solutions
we actually did find in our analysis. The result of this computation for
two and four family models is given in table \ref{tab_smest}.
This check clearly shows that our estimate gives the right order of
magnitude of the number of expected solutions, but the precise value might
differ by several standard deviations.
\begin{table}[htb!]
\begin{center}
\begin{tabular}{|r|r|r|r|}\hline
\# generations & \# of models found & estimated \# & suppression factor\\\hline
$2$ & $162921$ & $188908$ & $\approx 10^{-3}$\\
$3$ & $0$ & $0.2$ & $\approx 10^{-9}$\\
$4$ & $3898$ & $3310$ & $\approx 2\times10^{-5}$\\\hline
\end{tabular}
\caption{Comparison between the estimated number of solutions and the actual
number of solutions found for models with two, three and four generations.}
\label{tab_smest}
\end{center}
\end{table}

\section{Conclusions}
\label{SecCon}
In this work we have given an explicit statistical analysis of the vacuum structure
in a specific example of Type II orientifold models.
The bottom line is that a Standard Model-like
configuration with three families of quarks and leptons
in this class of models is statistically highly suppressed.
The concrete models which have so far been constructed in this setup
can therefore be regarded as exceptional within the vast majority of possible
solutions.
The same holds also for models with a Pati-Salam gauge group.

A statistical analysis of the  different observables
shows that there exist non-trivial correlations between some of them such as 
the rank of the gauge group and the chirality of the models.
Interestingly, these correlations hardly depend on whether we analyse them in the full set of solutions or only in the hidden sector of some
specific visible configuration. In that respect, the hidden sectors of a Standard Model or
Pati-Salam construction exhibit universal behavior and we expect this to hold for any other visible sector as long as the number of constraints it imposes is comparable.

Despite this general occurrence of correlations, we can quantitatively justify the hypothesis that the basic properties of Standard Model-like
constructions are to sufficiently good accuracy independent. Under this assumption,  we derive an estimate for the relative
frequency of three-generation Standard Models in our setup which is of the
order of $10^{-9}$. Of course, by requiring more of the 
phenomenological features of the Standard Model such as Yukawa couplings,
gauge couplings, soft supersymmetry breaking terms,  the occurrence of realistic models will get further
reduced. However, these properties really depend on the finer 
details of the models (see for instance \cite{cim03,Cvetic:2003ch,ao03,Kane:2004hm}) and an honest statistical treatment of them
is much harder to carry out.

An important question is how generic our results actually are.  Even though the available data for both frameworks are only partially compatible, we have performed a rough comparison with 
the statistics of the hidden sector of MSSM-like Gepner model orientifolds \cite{Dijkstra:2004ym} where possible. The geometrical properties of these small radius constructions differ considerably from those of the large radius regime. However, the results seem to suggest that the observables of the topological sector
exhibit quite similar distributions.  This is not so surprising as the diophantine structure of the consistency conditions describing this sector is essentially the same for both frameworks and since, after all, topological quantities should be protected against too drastic changes as one varies the coupling 
constants or radii. In fact, we regard this as strong hints towards a universal behaviour in the distribution of the topological observables also in other string constructions. 
By contrast, the distribution of the gauge couplings, which are dependent on the geometric moduli of the theory, do differ. It would be very desirable to collect more evidence supporting this picture. 
 
For this purpose it is important to study the statistics of 
other quite well understood classes of models like for instance heterotic string compactifications
on elliptically fibered Calabi-Yau manifolds (see for instance
\cite{Friedman:1997yq,Braun:2005ux,Blumenhagen:2005zg,Braun:2005xp}).
Questions in this direction include: 
 Does the distribution of the rank of the gauge group qualitatively
have the same shape? What is the abundance of Standard like models?
In addition, as we started to investigate in our first paper \cite{Blumenhagen:2004xx}, it would be interesting
to combine the T-dual picture of magnetised branes with additional three-form fluxes 
\cite{Blumenhagen:2003vr,Cascales:2003zp,Marchesano:2004xz,Gomis:2005wc} and
determine the effect on the statistics.

 \vskip 1cm
 {\noindent  {\Large \bf Acknowledgements}}
 \vskip 0.5cm 
\noindent
It is a pleasure to acknowledge interesting discussions with Bobby Acharya, Mirjam Cveti\v{c}, Frederik Denef,
Michael Douglas and Gary Shiu. 

We would like to thank the
Max-Planck-Rechenzentrum in Garching
and the system administrators at the Max-Planck-Institut f\"ur Physik
for technical support. 

 \vskip 2cm

\clearpage
\nocite{*}
\bibliography{paper}

\providecommand{\href}[2]{#2}\begingroup\raggedright\begin{thebibliography}{10}

\bibitem{Ibanez:1987sn}
L.~E. Ibanez, J.~E. Kim, H.~P. Nilles, and F.~Quevedo, ``Orbifold
  compactifications with three families of su(3) x su(2) x u(1)**n,'' {\em
  Phys. Lett.} {\bf B191} (1987)
282--286.

\bibitem{Casas:1988hb}
J.~A. Casas and C.~Munoz, ``Three generation su(3) x su(2) x u(1)-y models from
  orbifolds,'' {\em Phys. Lett.} {\bf B214} (1988)
63.

\bibitem{Braun:2005ux}
V.~Braun, Y.-H. He, B.~A. Ovrut, and T.~Pantev, ``A heterotic standard model,''
\href{http://www.arXiv.org/abs/hep-th/0501070}{{\tt hep-th/0501070}}.

\bibitem{Blumenhagen:2005mu}
R.~Blumenhagen, M.~Cvetic, P.~Langacker, and G.~Shiu, ``Toward realistic
  intersecting d-brane models,''
\href{http://www.arXiv.org/abs/hep-th/0502005}{{\tt hep-th/0502005}}.

\bibitem{bw98}
R.~Blumenhagen and A.~Wisskirchen, ``Spectra of 4d, n=1 type i string vacua on
  non-toroidal cy threefolds,'' {\em Phys. Lett.} {\bf B438} (1998) 52--60,
\href{http://www.arXiv.org/abs/hep-th/9806131}{{\tt hep-th/9806131}}.

\bibitem{Brunner:2004zd}
I.~Brunner, K.~Hori, K.~Hosomichi, and J.~Walcher, ``Orientifolds of gepner
  models,''
\href{http://www.arXiv.org/abs/hep-th/0401137}{{\tt hep-th/0401137}}.

\bibitem{bw04}
R.~Blumenhagen and T.~Weigand, ``Chiral supersymmetric gepner model
  orientifolds,'' {\em JHEP} {\bf 02} (2004) 041,
\href{http://www.arXiv.org/abs/hep-th/0401148}{{\tt hep-th/0401148}}.

\bibitem{Dijkstra:2004cc}
T.~P.~T. Dijkstra, L.~R. Huiszoon, and A.~N. Schellekens, ``Supersymmetric
  standard model spectra from rcft orientifolds,'' {\em Nucl. Phys.} {\bf B710}
  (2005) 3--57,
\href{http://www.arXiv.org/abs/hep-th/0411129}{{\tt hep-th/0411129}}.

\bibitem{Grana:2005jc}
M.~Grana, ``Flux compactifications in string theory: A comprehensive review,''
\href{http://www.arXiv.org/abs/hep-th/0509003}{{\tt hep-th/0509003}}.

\bibitem{Douglas:2003um}
M.~R. Douglas, ``The statistics of string / m theory vacua,'' {\em JHEP} {\bf
  05} (2003) 046,
\href{http://www.arXiv.org/abs/hep-th/0303194}{{\tt hep-th/0303194}}.

\bibitem{Ashok:2003gk}
S.~Ashok and M.~R. Douglas, ``Counting flux vacua,'' {\em JHEP} {\bf 01} (2004)
  060,
\href{http://www.arXiv.org/abs/hep-th/0307049}{{\tt hep-th/0307049}}.

\bibitem{Denef:2004ze}
F.~Denef and M.~R. Douglas, ``Distributions of flux vacua,'' {\em JHEP} {\bf
  05} (2004) 072,
\href{http://www.arXiv.org/abs/hep-th/0404116}{{\tt hep-th/0404116}}.

\bibitem{Giryavets:2004zr}
A.~Giryavets, S.~Kachru, and P.~K. Tripathy, ``On the taxonomy of flux vacua,''
  {\em JHEP} {\bf 08} (2004) 002,
\href{http://www.arXiv.org/abs/hep-th/0404243}{{\tt hep-th/0404243}}.

\bibitem{Dine:2004is}
M.~Dine, E.~Gorbatov, and S.~D. Thomas, ``Low energy supersymmetry from the
  landscape,''
\href{http://www.arXiv.org/abs/hep-th/0407043}{{\tt hep-th/0407043}}.

\bibitem{Misra:2004ky}
A.~Misra and A.~Nanda, ``Flux vacua statistics for two-parameter
  calabi-yau's,'' {\em Fortsch. Phys.} {\bf 53} (2005) 246--259,
\href{http://www.arXiv.org/abs/hep-th/0407252}{{\tt hep-th/0407252}}.

\bibitem{Conlon:2004ds}
J.~P. Conlon and F.~Quevedo, ``On the explicit construction and statistics of
  calabi-yau flux vacua,'' {\em JHEP} {\bf 10} (2004) 039,
\href{http://www.arXiv.org/abs/hep-th/0409215}{{\tt hep-th/0409215}}.

\bibitem{Denef:2004cf}
F.~Denef and M.~R. Douglas, ``Distributions of nonsupersymmetric flux vacua,''
  {\em JHEP} {\bf 03} (2005) 061,
\href{http://www.arXiv.org/abs/hep-th/0411183}{{\tt hep-th/0411183}}.

\bibitem{DeWolfe:2004ns}
O.~DeWolfe, A.~Giryavets, S.~Kachru, and W.~Taylor, ``Enumerating flux vacua
  with enhanced symmetries,'' {\em JHEP} {\bf 02} (2005) 037,
\href{http://www.arXiv.org/abs/hep-th/0411061}{{\tt hep-th/0411061}}.

\bibitem{Dienes:2004pi}
K.~R. Dienes, E.~Dudas, and T.~Gherghetta, ``A calculable toy model of the
  landscape,''
\href{http://www.arXiv.org/abs/hep-th/0412185}{{\tt hep-th/0412185}}.

\bibitem{Dine:2005yq}
M.~Dine, D.~O'Neil, and Z.~Sun, ``Branches of the landscape,'' {\em JHEP} {\bf
  07} (2005) 014,
\href{http://www.arXiv.org/abs/hep-th/0501214}{{\tt hep-th/0501214}}.

\bibitem{Acharya:2005ez}
B.~S. Acharya, F.~Denef, and R.~Valandro, ``Statistics of m theory vacua,''
\href{http://www.arXiv.org/abs/hep-th/0502060}{{\tt hep-th/0502060}}.

\bibitem{Distler:2005hi}
J.~Distler and U.~Varadarajan, ``Random polynomials and the friendly
  landscape,''
\href{http://www.arXiv.org/abs/hep-th/0507090}{{\tt hep-th/0507090}}.

\bibitem{Douglas:2005hq}
M.~R. Douglas and Z.~Lu, ``Finiteness of volume of moduli spaces,''
\href{http://www.arXiv.org/abs/hep-th/0509224}{{\tt hep-th/0509224}}.

\bibitem{Banks:2003es}
T.~Banks, M.~Dine, and E.~Gorbatov, ``Is there a string theory landscape?,''
  {\em JHEP} {\bf 08} (2004) 058,
\href{http://www.arXiv.org/abs/hep-th/0309170}{{\tt hep-th/0309170}}.

\bibitem{Banks:2004xh}
T.~Banks, ``Landskepticism or why effective potentials don't count string
  models,''
\href{http://www.arXiv.org/abs/hep-th/0412129}{{\tt hep-th/0412129}}.

\bibitem{Dijkstra:2004ym}
T.~P.~T. Dijkstra, L.~R. Huiszoon, and A.~N. Schellekens, ``Chiral
  supersymmetric standard model spectra from orientifolds of gepner models,''
  {\em Phys. Lett.} {\bf B609} (2005) 408--417,
\href{http://www.arXiv.org/abs/hep-th/0403196}{{\tt hep-th/0403196}}.

\bibitem{Kumar:2004pv}
J.~Kumar and J.~D. Wells, ``Landscape cartography: A coarse survey of gauge
  group rank and stabilization of the proton,'' {\em Phys. Rev.} {\bf D71}
  (2005) 026009,
\href{http://www.arXiv.org/abs/hep-th/0409218}{{\tt hep-th/0409218}}.

\bibitem{Blumenhagen:2004xx}
R.~Blumenhagen, F.~Gmeiner, G.~Honecker, D.~L{\"u}st, and T.~Weigand, ``The
  statistics of supersymmetric d-brane models,'' {\em Nucl. Phys.} {\bf B713}
  (2005) 83--135,
\href{http://www.arXiv.org/abs/hep-th/0411173}{{\tt hep-th/0411173}}.

\bibitem{Kumar:2005hf}
J.~Kumar and J.~D. Wells, ``Surveying standard model flux vacua on t**6/z(2) x
  z(2),''
\href{http://www.arXiv.org/abs/hep-th/0506252}{{\tt hep-th/0506252}}.

\bibitem{Arkani-Hamed:2005yv}
N.~Arkani-Hamed, S.~Dimopoulos, and S.~Kachru, ``Predictive landscapes and new
  physics at a tev,''
\href{http://www.arXiv.org/abs/hep-th/0501082}{{\tt hep-th/0501082}}.

\bibitem{Vafa:2005ui}
C.~Vafa, ``The string landscape and the swampland,''
\href{http://www.arXiv.org/abs/hep-th/0509212}{{\tt hep-th/0509212}}.

\bibitem{ciu03}
P.~G. C\'amara, L.~E. Ib\'a\~nez, and A.~M. Uranga, ``Flux-induced
  susy-breaking soft terms,'' {\em Nucl. Phys.} {\bf B689} (2004) 195--242,
\href{http://www.arXiv.org/abs/hep-th/0311241}{{\tt hep-th/0311241}}.

\bibitem{ggjl03}
M.~Grana, T.~W. Grimm, H.~Jockers, and J.~Louis, ``Soft supersymmetry breaking
  in calabi-yau orientifolds with d-branes and fluxes,'' {\em Nucl. Phys.} {\bf
  B690} (2004) 21--61,
\href{http://www.arXiv.org/abs/hep-th/0312232}{{\tt hep-th/0312232}}.

\bibitem{Lust:2004fi}
D.~L{\"u}st, S.~Reffert, and S.~Stieberger, ``Flux-induced soft supersymmetry
  breaking in chiral type iib orientifolds with d3/d7-branes,'' {\em Nucl.
  Phys.} {\bf B706} (2005) 3--52,
\href{http://www.arXiv.org/abs/hep-th/0406092}{{\tt hep-th/0406092}}.

\bibitem{lrs04a}
D.~L{\"u}st, S.~Reffert, and S.~Stieberger, ``Mssm with soft susy breaking
  terms from d7-branes with fluxes,''
\href{http://www.arXiv.org/abs/hep-th/0410074}{{\tt hep-th/0410074}}.

\bibitem{Lust:2005bd}
D.~L{\"u}st, P.~Mayr, S.~Reffert, and S.~Stieberger, ``F-theory flux,
  destabilization of orientifolds and soft terms on d7-branes,''
\href{http://www.arXiv.org/abs/hep-th/0501139}{{\tt hep-th/0501139}}.

\bibitem{Forste:2000hx}
S.~F{\"o}rste, G.~Honecker, and R.~Schreyer, ``Supersymmetric z(n) x z(m)
  orientifolds in 4d with d-branes at angles,'' {\em Nucl. Phys.} {\bf B593}
  (2001) 127--154,
\href{http://www.arXiv.org/abs/hep-th/0008250}{{\tt hep-th/0008250}}.

\bibitem{Cvetic:2001nr}
M.~Cvetic, G.~Shiu, and A.~M. Uranga, ``Chiral four-dimensional n = 1
  supersymmetric type iia orientifolds from intersecting d6-branes,'' {\em
  Nucl. Phys.} {\bf B615} (2001) 3--32,
\href{http://www.arXiv.org/abs/hep-th/0107166}{{\tt hep-th/0107166}}.

\bibitem{Larosa:2003mz}
M.~Larosa and G.~Pradisi, ``Magnetized four-dimensional z(2) x z(2)
  orientifolds,'' {\em Nucl. Phys.} {\bf B667} (2003) 261--309,
\href{http://www.arXiv.org/abs/hep-th/0305224}{{\tt hep-th/0305224}}.

\bibitem{Dudas:2005jx}
E.~Dudas and C.~Timirgaziu, ``Internal magnetic fields and supersymmetry in
  orientifolds,'' {\em Nucl. Phys.} {\bf B716} (2005) 65--87,
\href{http://www.arXiv.org/abs/hep-th/0502085}{{\tt hep-th/0502085}}.

\bibitem{Blumenhagen:2005tn}
R.~Blumenhagen, M.~Cvetic, F.~Marchesano, and G.~Shiu, ``Chiral d-brane models
  with frozen open string moduli,'' {\em JHEP} {\bf 03} (2005) 050,
\href{http://www.arXiv.org/abs/hep-th/0502095}{{\tt hep-th/0502095}}.

\bibitem{Uranga:2000xp}
A.~M. Uranga, ``D-brane probes, rr tadpole cancellation and k-theory charge,''
  {\em Nucl. Phys.} {\bf B598} (2001) 225--246,
\href{http://www.arXiv.org/abs/hep-th/0011048}{{\tt hep-th/0011048}}.

\bibitem{Ibanez:2001nd}
L.~E. Ibanez, F.~Marchesano, and R.~Rabadan, ``Getting just the standard model
  at intersecting branes,'' {\em JHEP} {\bf 11} (2001) 002,
\href{http://www.arXiv.org/abs/hep-th/0105155}{{\tt hep-th/0105155}}.

\bibitem{Witten:1982fp}
E.~Witten, ``An su(2) anomaly,'' {\em Phys. Lett.} {\bf B117} (1982)
324--328.

\bibitem{Antoniadis:2000en}
I.~Antoniadis, E.~Kiritsis, and T.~N. Tomaras, ``A d-brane alternative to
  unification,'' {\em Phys. Lett.} {\bf B486} (2000) 186--193,
\href{http://www.arXiv.org/abs/hep-ph/0004214}{{\tt hep-ph/0004214}}.

\bibitem{Blumenhagen:2001te}
R.~Blumenhagen, B.~K{\"o}rs, D.~L{\"u}st, and T.~Ott, ``The standard model from
  stable intersecting brane world orbifolds,'' {\em Nucl. Phys.} {\bf B616}
  (2001) 3--33,
\href{http://www.arXiv.org/abs/hep-th/0107138}{{\tt hep-th/0107138}}.

\bibitem{gareyjohnson1979}
M.~R. Garey and D.~S. Johnson, {\em Computers and Intractability, a Guide to
  the Theory of NP-Completeness}.
\newblock Freeman San Francisco, 1979.

\bibitem{Gato-Rivera:2005qd}
B.~Gato-Rivera and A.~N. Schellekens, ``Remarks on global anomalies in rcft
  orientifolds,''
\href{http://www.arXiv.org/abs/hep-th/0510074}{{\tt hep-th/0510074}}.

\bibitem{Marchesano:2004xz}
F.~Marchesano and G.~Shiu, ``Building mssm flux vacua,'' {\em JHEP} {\bf 11}
  (2004) 041,
\href{http://www.arXiv.org/abs/hep-th/0409132}{{\tt hep-th/0409132}}.

\bibitem{Cvetic:2005bn}
M.~Cvetic, T.~Li, and T.~Liu, ``Standard-like models as type iib flux vacua,''
  {\em Phys. Rev.} {\bf D71} (2005) 106008,
\href{http://www.arXiv.org/abs/hep-th/0501041}{{\tt hep-th/0501041}}.

\bibitem{Blumenhagen:2003jy}
R.~Blumenhagen, D.~L{\"u}st, and S.~Stieberger, ``Gauge unification in
  supersymmetric intersecting brane worlds,'' {\em JHEP} {\bf 07} (2003) 036,
\href{http://www.arXiv.org/abs/hep-th/0305146}{{\tt hep-th/0305146}}.

\bibitem{cim03}
D.~Cremades, L.~E. Ib\'a\~nez, and F.~Marchesano, ``Yukawa couplings in
  intersecting d-brane models,'' {\em JHEP} {\bf 07} (2003) 038,
\href{http://www.arXiv.org/abs/hep-th/0302105}{{\tt hep-th/0302105}}.

\bibitem{Cvetic:2003ch}
M.~Cvetic and I.~Papadimitriou, ``Conformal field theory couplings for
  intersecting d-branes on orientifolds,'' {\em Phys. Rev.} {\bf D68} (2003)
  046001,
\href{http://www.arXiv.org/abs/hep-th/0303083}{{\tt hep-th/0303083}}.

\bibitem{ao03}
S.~A. Abel and A.~W. Owen, ``Interactions in intersecting brane models,'' {\em
  Nucl. Phys.} {\bf B663} (2003) 197--214,
\href{http://www.arXiv.org/abs/hep-th/0303124}{{\tt hep-th/0303124}}.

\bibitem{Kane:2004hm}
G.~L. Kane, P.~Kumar, J.~D. Lykken, and T.~T. Wang, ``Some phenomenology of
  intersecting d-brane models,'' {\em Phys. Rev.} {\bf D71} (2005) 115017,
\href{http://www.arXiv.org/abs/hep-ph/0411125}{{\tt hep-ph/0411125}}.

\bibitem{Friedman:1997yq}
R.~Friedman, J.~Morgan, and E.~Witten, ``Vector bundles and f theory,'' {\em
  Commun. Math. Phys.} {\bf 187} (1997) 679--743,
\href{http://www.arXiv.org/abs/hep-th/9701162}{{\tt hep-th/9701162}}.

\bibitem{Blumenhagen:2005zg}
R.~Blumenhagen, G.~Honecker, and T.~Weigand, ``Non-abelian brane worlds: The
  heterotic string story,''
\href{http://www.arXiv.org/abs/hep-th/0510049}{{\tt hep-th/0510049}}.

\bibitem{Braun:2005xp}
V.~Braun, Y.-H. He, B.~A. Ovrut, and T.~Pantev, ``Moduli dependent mu-terms in
  a heterotic standard model,''
\href{http://www.arXiv.org/abs/hep-th/0510142}{{\tt hep-th/0510142}}.

\bibitem{Blumenhagen:2003vr}
R.~Blumenhagen, D.~L{\"u}st, and T.~R. Taylor, ``Moduli stabilization in chiral
  type iib orientifold models with fluxes,'' {\em Nucl. Phys.} {\bf B663}
  (2003) 319--342,
\href{http://www.arXiv.org/abs/hep-th/0303016}{{\tt hep-th/0303016}}.

\bibitem{Cascales:2003zp}
J.~F.~G. Cascales and A.~M. Uranga, ``Chiral 4d n = 1 string vacua with
  d-branes and nsns and rr fluxes,'' {\em JHEP} {\bf 05} (2003) 011,
\href{http://www.arXiv.org/abs/hep-th/0303024}{{\tt hep-th/0303024}}.

\bibitem{Gomis:2005wc}
J.~Gomis, F.~Marchesano, and D.~Mateos, ``An open string landscape,''
\href{http://www.arXiv.org/abs/hep-th/0506179}{{\tt hep-th/0506179}}.

\end{thebibliography}\endgroup
\bibliographystyle{utphys}

\end{document}